\documentclass{amsart}
\usepackage{amsmath}
\usepackage{amscd}
\usepackage[mathscr]{eucal}
\usepackage{amssymb} 
\usepackage{latexsym}
\usepackage{amsthm}
\usepackage{verbatim}

\title[elementary integral calculus on superspace ${\mathfrak{R}}^{m|n}$]
{Remarks\\ on elementary integral calculus\\ for 
supersmooth functions on superspace ${\mathfrak{R}}^{m|n}$}
\author{Atsushi Inoue} 
\date{\today}
\address{Emeritus, Tokyo Institute of Technology}
\address{Department of mathematics, Faculty of sciences, Tokyo Institute of Technology}
 %46S60}
\dedicatory{Dedicated to the memory of late Professor Seizo ITO}
\email{inoue@math.titech.ac.jp} %atlom-inoue60@nifty.com.}

\def\where{\quad\text{where}\quad}

\def\with{\quad\text{with}\quad}
\def\for{\quad\text{for}\quad}

\def\et{\quad\text{and}\quad}

\def\ess.sup{{\operatornamewithlimits{ess.sup}}}
\def\sdet{\operatorname{sdet}}

\def\sgn{\operatorname{sgn\,}}

\def\euc{{\mathbb R}}

\def\eucm{{\mathbb R}^m}

\def\fR{{\mathfrak R}}
\def\fC{{\mathfrak C}}
\def\CSS{{\mathcal C}_{{\mathrm{SS}}}}

\def\ccsl{{/\kern-0.5em {\mathcal C}}} %{{/\kern-0.5em {\mathcal C}}}
\def\cbsl{{/\kern-0.65em {\mathcal B}}} 
\def\cssl{{/\kern-0.55em {\mathcal S}}} 
\def\cdsl{{/\kern-0.7em {\mathcal D}}} 
\def\cesl{{/\kern-0.58em {\mathcal E}}} 
\def\clsl{{/\kern-0.6em {\mathcal L}}} 
\def\chsl{{/\kern-0.65em {\mathcal H}}} 
\def\cpsl{{/\kern-0.6em {\mathcal P}}} 
\def\comsl{{/\kern-0.6em {\mathcal O}}_{M}}
\def\superon{{{\mathfrak R}{}^{0|n}}}
\def\supermn{{{\mathfrak R}{}^{m|n}}}

\def\supermo{{{\mathfrak R}{}^{m|0}}}

\def\dirac{/\kern-0.7em D}
\def\cev{{\fC}_{\mathrm{ev}}}
\def\cod{{\fC}_{\mathrm{od}}}
\def\rev{{\fR}_{\mathrm{ev}}}
\def\rod{{\fR}_{\mathrm{od}}}
\def\dt{\frac{d}{dt}}

\def\Ber{\operatorname{Ber}} 

\def\Biint{{\mathfrak{B}}\!\!\!-\!\!\!\!{\int{\!\!\!\!\!\!}\int}}  %{\mathcal{B}}\!\!\!\!\!\!{\int{\!\!\!\!\!}\int}} %{{B}\!\!\!-\!\!\!\!\iint}  %{\mathfrak{B}\!\!\!\!\!\!\!{\int\!\!\!\int}}
\def\VViint{{\mathfrak{V}\!\mathfrak{V}}\!\!\!-\!\!\!\!{\int{\!\!\!\!\!}\int}} %\!\!\!\!\!\!\!\!{\int{\!\!\!\!\!}\int}}  %{{V\!V}\!\!\!-\!\!\!\!\iint}%{\operatorname{V\!V}\!\!\!\!\!\!\!\!\!\!{\int\!\!\!\int}}
\numberwithin{equation}{section}
\newtheorem{theorem}{Theorem}[section]
\newtheorem{lemma}[theorem]{Lemma}
\newtheorem{proposition}[theorem]{Proposition}
\newtheorem{corollary}[theorem]{Corollary}
\newtheorem{remark}[theorem]{Remark} %[section]
\newtheorem{definition}{Definition}[section]
\newtheorem{assumption}{Assumption}
\newtheorem{example}{Example}[section] 
\newtheorem{claim}{Claim}[section]

\begin{document}

\baselineskip=16pt
\vspace{-0.7cm}
\begin{abstract}
After introducing Berezin integral for polynomials of odd variables, we 
develop the elementary integral calculus based on supersmooth functions on the superspace ${\mathfrak{R}}^{m|n}$.
Here, ${\mathfrak{R}}$ is the Fr\'echet-Grassmann algebra with countably infinite Grassmann generators, which plays 
the role of real number field ${\mathbb{R}}$.
As is well-known that the formula of change of variables under integral sign is indispensable not only to treat PDE 
applying funtional analytic method but also to introduce analysis on supermanifolds. 
But, if we define naively the integral for supersmooth functions, there exists discrepancy which should be ameliorated.
Here, we extend the contour integral modifying the parameter space introduced basically by de Witt, Rogers and Vladimirov and Volovich.
%We reconsider the new volume form introduced by Rothstein and compare it with other works.
\end{abstract}
\maketitle

\section{Introduction: Problem and Results}
\subsection{Problem}
How to get the Feynman-like representation for the fundamental solution of  the Dirac equation 
with external electro-magnetic potentials?  (See, Inoue~\cite{ino98-2} for the free Dirac equation or Inoue~\cite{ino02} for the Weyl equation with external electro-magnetic field.)
To answer this Feynman's question affirmatively but also to offer a prototype of new procedures to study other systems of PDE without diagonalizing coefficient matrices, we need to construct not only differential but also integral calculus based on a non-commutative algebra with countably many Grassmann generators.

Though, there are so many papers concerning elementary calculus prefixed ``super-'' which is based 
on Banach space, such as ${\mathfrak{B}}_{L}$ or ${\mathfrak{B}}(={\mathfrak{B}}_{\infty})$, 
there is rather few dealing fully with elementary integral calculus
based on Fr\'echet-Grassmann algebra, such as ${\fR}$ introduced in Inoue and Maeda~\cite{IM91}, Inoue~\cite{ino92-1, ino09}. 
But a part of the elementary differential calclulus based on ${\fR}$ is mentioned, for example, in P.Bryant~\cite{bry87}, Y.Choquet-Bruhat~\cite{ChB90}, S. Matsumoto and K. Kakazu~\cite{MK}, K. Yagi~\cite{yag88}.
By the way, for the elementary differential calculus on a general Fr\'echet space, Hamilton's work~\cite{ham82} is transparent.

In our previous paper \cite{ino09},  we introduce and characterize the so-called supersmooth functions (alias superfield) 
on ${\fR}^{m|n}$.
In order to treat certain systems of PDE without diagonalization, we regard matrices as differential operators acting on supersmooth functions and to apply method of functional analysis to that PDE,  we need to develop integral calculus on ${\fR}^{m|n}$ which admit the formula of the change of variables under integral sign.
Applying this integration theory, we may construct a parametrix having the representation of Fourier integral operator type with the phase function (roughly saying, with the matrix valued-phase function) satisfying the Hamilton-Jacobi equation. 
It should be remarked that since $\fR$ is an infinite dimensional Fr\'echet space, we need a care.

\begin{remark}
Matrices appeared are confined to $2^d{\times} 2^d$-type, because we use Clifford algebras to expand matrices which have differential operator representations on Grassmann algebras.
To treat matrices such as $3\times 3$, it seems necessary to develop 
another non-commutative space and analysis on it. See for example, Martin~\cite{mar59-1} and one may get some hints from Khare~\cite{kha}, or Campoamor-Stursberg, Rausch de Traubenberg~\cite{CS-RT}.
\end{remark}

Concerning supersmooth functions only with even variables, we have the theory resembling to the integral in complex analysis, see, de~Witt \cite{DW2} and Rogers \cite{rog80}.
On the other hand, functions only with odd variables, we have the well-known Berezin integral.

In order to treat even and odd variables on equal footing, we need to mix naturally these variables, for example, the supersymmetric transformations are generated by mixing both variables.
Therefore, we need to construct integration theory
which admits the wide class of the change of variables under integral sign.

As is well-known, to study a scalar PDE by applying functional analysis, 
we use essentially the following tools: Taylor expansion, 
integration by parts, the formula for the change of variables under integral sign and Fourier transformation.
Therefore, beside the elementary differential calculus, it is necessary to develop the elementary integral calculus on superspace ${\fR}^{m|n}$, both consist of the elementary superanalysis.
But as is explained soon later, after defining differentials $dx_j$ and $d\theta_k$ properly, we have the relations
$$
\begin{cases}
dx_j\wedge dx_k=-dx_k\wedge dx_j&\mbox{for even variables $\{x_j\}_{j=1}^m$},\\
d\theta_j\wedge d\theta_k=d\theta_k\wedge d\theta_j &\mbox{for odd variables $\{\theta_k\}_{k=1}^n$, which differs from ordinary one}.
\end{cases}
$$
Therefore, the integration containing odd variables doesn't necessarily follow from our conventional intuition.

\begin{remark}
It is rather straight forward to extend the notion defined on Euclidian space to that on Banach space, but not so on Fre\'chet space. 
For example, the implicit function theorem is the typical one which is not extendable to general infinite dimensional Fr\'echet spaces without additional conditions. Therefore, we need a care to change the space from Roger's ${\mathfrak{B}}$ to our ${\fR}$.
Moreover, we have another algebraic operation, called the multiplication, in Fre\'chet-Grassmann or Banach-Grassmann algebras. Like the Fr\'echet differentiability on ${\mathbb{C}}$(=not only isomorphic to $\euc^2$ but also have the multiplication) leads naturally to the notion of analyticity, the multiplication in those algebras yields the new notion called supersmoothness (see, \cite{ino09}).
\end{remark}

\begin{definition}\label{def3-1.1}
For a set $U\subset\euc^m$, we define $\pi_{\mathrm{B}}^{-1}(U)=\{X\in \supermo\;|\; \pi_{\mathrm{B}}(X)\in U\}$.
A set ${\mathfrak{U}}_{\mathrm{ev}}\subset\supermo$ 
is called an ``even superdomain'' 
if $U=\pi_{\mathrm B}({\mathfrak{U}}_{\mathrm{ev}})
\subset {\mathbb R}^m$ is open, connected and
$\pi_{\mathrm B}^{-1}(U)={\mathfrak{U}}_{\mathrm{ev}}$.
$U$ is denoted also by ${\mathfrak{U}}_{\mathrm{ev},\mathrm B}$.
When $ {\mathfrak{U}} \subset \supermn$ is represented by
${\mathfrak{U}}={\mathfrak{U}}_{\mathrm{ev}} \times {\fR}_{\mathrm{od}}^n$ with an even superdomain 
${\mathfrak{U}}_{\mathrm{ev}}\subset\supermo$,
${\mathfrak{U}}$ is called a ``superdomain'' in ${\fR}^{m|n}$.
\end{definition}

\begin{definition}[A naive definition of Berezin integral]\label{NDI}
For a super domain ${\mathfrak{U}}={\mathfrak{U}}_{\mathrm{ev}} \times {{\fR}^{0|n}}$
%with ${\mathfrak{U}}_{\mathrm{ev}}\subset\rev^m={\fR}^{m|0}$ 
and a supersmooth function $u(x,\theta)=\sum_{|a|\le n}\theta^a u_a(x):{\mathfrak{U}}\to{\fR}$, we ``define'' its integral as
\begin{equation}
\begin{gathered}
\Biint_{\!\!\!\mathfrak{U}} dx d\theta\, u(x,\theta)=\int_{{\mathfrak{U}}_{\mathrm{ev}}}dx\bigg(\int_{{\fR}^{0|n}}d\theta\, u(x,\theta)\bigg)
=\int_{\pi_{\mathrm{B}}({\mathfrak{U}}_{\mathrm{ev}})}dq\,u_{\bar{1}}(q),\\
\where
\int_{{\fR}^{0|n}}d\theta\, u(x,\theta)=\frac{\partial}{\partial\theta_n}{\cdots}
\frac{\partial}{\partial\theta_1}u(x,\theta)\bigg|_{\theta_1={\cdots}=\theta_n=0}=u_{\bar{1}}(x)\et \bar{1}=(\overbrace{1,{\cdots},1}^{n}).
\end{gathered}
\label{BF}
\end{equation}
In the above, $u_{\bar{1}}(x)$ is the Grassmann continuation of $u_{\bar{1}}(q)$.
\end{definition}
Desiring that the standard formula of \underline{the change of variables under integral sign} (={\bf {CVF}}) holds by replacing standard Jacobian with super Jacobian(= super determinant of Jacobian matrix) on ${\fR}^{m|n}$, we have 
%the following \underline{if an integrand has compact support}:
\begin{theorem}\label{B1}
Let ${\mathfrak{U}}={\mathfrak{U}}_{\mathrm{ev}} \times {{\fR}^{0|n}}\subset{\fR}^{m|n}_X$ and 
${\mathfrak{V}}={\mathfrak{V}}_{\mathrm{ev}} \times {{\fR}^{0|n}}\subset{\fR}^{m|n}_Y$ be given. Let
\begin{equation}
\varphi:{\mathfrak{V}}\ni Y=(y,\omega)\to X=(x,\theta)=(\varphi_{\bar0}(y,\omega),\varphi_{\bar1}(y,\omega))\in{\mathfrak{U}}
\label{4-17}
\end{equation}
be a {supersmooth} diffeomorphism from ${\mathfrak{V}}$ onto ${\mathfrak{U}}$,
that is, 
\begin{equation}
{\sdet} J(\varphi)(y,\omega)\neq0\et\varphi({\mathfrak{V}})={\mathfrak{U}}\where
J(\varphi)(y,\omega)=
\begin{pmatrix}
\frac {\partial \varphi_{\bar0}(y,\omega)}{\partial y}&\frac {\partial\varphi_{\bar1}(y,\omega)}{\partial y}\\[2ex]
\frac {\partial \varphi_{\bar0}(y,\omega)}{\partial \omega}&\frac {\partial\varphi_{\bar1}(y,\omega)}{\partial \omega}
\end{pmatrix}.
\label{4-17-1}
\end{equation}
Then, for any function $u\in{\CSS}({\mathfrak{U}}:{\fC})$ with \underline{compact support}, 
that is, $u(x,\theta)=\sum_{|a|\le n}\theta^a u_a(x)$ where
$u_a(x_{\mathrm{B}})\in C^{\infty}_0({\mathfrak{U}}_{\mathrm{ev, B}}:{\fR})$ for all $a\in\{0,1\}^n$ except $a=\bar{1}$,
we have CVF %the change of variables formula
\begin{equation}
\Biint_{\!\!\!\mathfrak{U}}dxd\theta \, u(x,\theta)
=\Biint_{\!\!\!\varphi^{-1}(\mathfrak{U})}dyd\omega \,{\sdet} J(\varphi)(y,\omega) 
u(\varphi(y,\omega)).
\label{B2.2.12}
\end{equation}
\end{theorem}

\begin{remark}
Decomposing an even supermatrix $M$ given by
$$
M=\begin{pmatrix}
A&C\\
D&B
\end{pmatrix}
\with
{\left\{
\begin{gathered}
A=(a_{ij}), \; B=(b_{k\ell}),\;\; a_{ij},\; b_{k\ell}\in\rev,\\
C=(c_{i\ell}), \; D=(d_{kj}),\;\;  c_{i\ell},\; d_{kj}\in\rod,
\end{gathered}
\right.},
$$
we put
$$
\sdet M=\begin{cases}
\det A{\cdot}\det(B-DA^{-1}C)^{-1}&\mbox{if $\det A_{\mathrm{B}}\neq0$ where $A_{\mathrm{B}}=(\pi_{\mathrm{B}}a_{ij})$},\\
\det(A-CB^{-1}D){\cdot}\det B^{-1}&\mbox{if $\det B_{\mathrm{B}}\neq0$ where $B_{\mathrm{B}}=(\pi_{\mathrm{B}}b_{k\ell})$}.
\end{cases}
$$
\end{remark}

\begin{remark}
Seemingly, this theorem implies that
\underline{Berezin ``measure'' $D_0(x,\theta)$} is transformed by $\varphi$ as
\begin{equation}
(\varphi^*D_0(x,\theta))(y,\omega)=D_0(y,\omega){\cdot}{\sdet}J(\varphi)(y,\omega),\\
\label{Ber-mea}
\end{equation}
where
$$
D_0(x,\theta)=dx_1\wedge{\cdots}\wedge dx_m\otimes
\frac{\partial}{\partial\theta_n}{\cdots}\frac{\partial}{\partial\theta_1}
=dx_1{\cdots}dx_m {\cdot}{\partial}_{\theta_n}{\cdots}{\partial}_{\theta_1}
=dx\partial_{\theta}^{\bar1},\; D_0(y,\omega)=dy\partial_{\omega}^{\bar1}.
$$
But this assertion is shown to be false in general by the following example.
Moreover, we remark also that the condition of ``the compact supportness of integrands'' above, seems not only
cumbersome from conventional point of view but also fatal in holomorphic category.
\end{remark}

\begin{example}\label{Example1} 
Let
${\mathfrak{U}}=\pi_{\mathrm B}^{-1}({\Omega})\times\rod^2\subset {\fR}^{1|2}$ 
with ${\Omega}=(0,1)$, $\pi_{\mathrm{B}}:{\fR}^{1|0}\to  \euc$ and let $u$ be supersmooth on ${\fR}^{1|2}$ with value in ${\fR}$ such that
$u(x,\theta)=u_{\bar{0}}(x)+\theta_1\theta_2u_{\bar{1}}(x)$.
Then, we have
$$
\Biint_{\!\!\!\mathfrak{U}} dx d\theta\,u(x,\theta)=
\int_{\Omega} dx\int d\theta\, u(x,\theta)=\int_{\pi_{\mathrm B}^{-1}({\Omega})}dx\,u_{\bar{1}}(x)
=\int_{\Omega}dq\,u_{\bar{1}}(q).
$$
But, if we use the coordinate change
\begin{equation}
\varphi:(y,\omega)\to (x,\theta)\with
x=y+\omega_1\omega_2\phi(y),\;\theta_k=\omega_k: {\mathfrak{U}}\to {\mathfrak{U}}
\label{ex-cv}
\end{equation}
whose Berezinian is
$$
{\Ber}(\varphi)(y,\omega)={\sdet}J(\varphi)(y,\omega)=1+\omega_1\omega_2\phi'(y)
\where
J(\varphi)(y,\omega)=\begin{pmatrix}
1+\omega_1\omega_2\phi'(y)&0&0\\
\omega_2\phi(y)&1&0\\
-\omega_1\phi(y)&0&1
\end{pmatrix},
$$
and if we assume that the formula \eqref{B2.2.12} holds,
then since
$$
\begin{gathered}
u(\varphi(y,\omega))=u_{\bar{0}}(y+\omega_1\omega_2\phi(y))+\omega_1\omega_2u_{\bar{1}}(y+\omega_1\omega_2\phi(y))
=u_{\bar{0}}(y)+\omega_1\omega_2(\phi(y)u_{\bar{0}}'(y)+u_{\bar{1}}(y)),\\
\et 
(1+\omega_1\omega_2\phi'(y))u(\varphi(y,\omega))=u_{\bar{0}}(y)+\omega_1\omega_2(\phi(y)u_{\bar{0}}'(y)+\phi'(y)u_{\bar{0}}(y)+u_{\bar{1}}(y)),
\end{gathered}
$$
we have
$$
\Biint_{\!\!\!\varphi^{-1}(\mathfrak{U})} dy d\omega\,(1+\omega_1\omega_2\phi'(y))u(\varphi(y,\omega))
=\int_{\pi_{\mathrm B}^{-1}({\Omega})}dy\,(\phi(y)u_{\bar{0}}(y))'+ \int_{\pi_{\mathrm B}^{-1}({\Omega})}dx\,u_{\bar{1}}(x).
$$
Therefore, if $\displaystyle{\int_{\pi_{\mathrm B}^{-1}({\Omega})}}dy\,(\phi(y)u_{\bar{0}}(y))'\neq 0$, then
$\displaystyle{{\int{\!\!\!\!\!}\int}_{\mathfrak{U}}} D_0(x,\theta)\,u(x,\theta)\neq \displaystyle{{\int{\!\!\!\!\!}\int}_{\varphi^{-1}(\mathfrak{U})}} D_0(y,\omega)\,u(\varphi(y,\omega))$. 
This implies that if we apply \eqref{BF} as definition, 
the change of variables formula doesn't hold when, for example, the integrand hasn't compact support.
\end{example}

\subsection{Results}
Though there seems several methods (for example, due to Rothstein~\cite{roth87}, Zirnbauer~\cite{zir96}) to remedy such inconsistency in case ${\mathfrak{B}}_{L}^{m|n}$ or ${\mathfrak{B}}^{m|n}$,  we
modify Vladimirov and Volovich~\cite{VV84} to get our results.

\begin{definition}[Parameter set, paths and integral]\label{4.9}
We prepare a domain $\Omega$ in ${\euc}^m$ and put $\widetilde{\Omega}=\Omega\times\rod^n$, called a parameter set.
\par
(1)%{}_{\scriptstyle{}\atop\scriptstyle{}}
Let $\gamma\in C^\infty(\widetilde{\Omega}:{\fR}^{m|n})$ with 
$\gamma(q,{\vartheta})=(\gamma_{\bar0}(q,{\vartheta}), \gamma_{\bar1}(q,{\vartheta}))=(\gamma_{\bar0,j}(q,{\vartheta}), \gamma_{\bar1,k}(q,{\vartheta}))_{\scriptstyle{j=1,{\cdots},m}\atop\scriptstyle{k=1,{\cdots},n}}$ be given such that  
$$
\gamma_{\bar0,j}(q,{\vartheta})=\sum_{|a|\le n}{\vartheta}^a\gamma_{\bar0,j,a}(q)\in{\rev},\quad
\gamma_{\bar1,k}(q,{\vartheta})=\sum_{|a|\le n}{\vartheta}^a\gamma_{\bar1,k,a}(q)\in{\rod}
$$
where
$$
\begin{gathered}
\gamma_{\bar0,j,a}(q)=\sum_{|{\bf{I}}|=|a|({\mathrm{mod}}2)} \gamma_{\bar0,j,a,{\bf{I}}}(q)\sigma^{\bf{I}},\;\;
\gamma_{\bar1,k,a}(q)=\sum_{|{\bf{J}}|=|a|+1({\mathrm{mod}}2)} \gamma_{\bar1,k,a,{\bf{J}}}(q)\sigma^{\bf{J}} %,\\
\end{gathered}
$$
with
$$
\gamma_{\bar0,a,{\bf{I}}}(q),\; \gamma_{\bar1,a,{\bf{J}}}(q)\in C^\infty(\Omega:{\mathbb C}^n) \et
\gamma_{\bar0,\bar{0},\tilde{0}}(q)\in C^\infty(\Omega:{\mathbb R}^m),\; \bar{0}=(0,{\cdots},0),\; \tilde{0}=(0,{\cdots}).
$$
In case
$$
\sdet J(\gamma)(q,{\vartheta})\neq 0  \where
J(\gamma)(q,{\vartheta})=\frac{\partial\gamma(q,{\vartheta})}{\partial(q,{\vartheta})}=
\begin{pmatrix}
\frac{\partial\gamma_{\bar0}(q,{\vartheta})}{\partial q}&\frac{\partial\gamma_{\bar1}(q,{\vartheta})}{\partial q}\\
\frac{\partial\gamma_{\bar0}(q,{\vartheta})}{\partial{\vartheta}}&\frac{\partial\gamma_{\bar1}(q,{\vartheta})}{\partial{\vartheta}}
\end{pmatrix}, 
$$
we call this $\gamma$ as a ``path''  from $\widetilde{\Omega}$ into ${\fR}^{m|n}$ and its image is called 
a \underline{foliated singular manifold}:
$$
{\mathfrak{M}}={\mathfrak{M}}(\gamma,\Omega)=\gamma(\widetilde{\Omega})=\{(x,\theta)\in \supermn\;\big|\;
x=\gamma_{\bar0}(q,{\vartheta}), \theta=\gamma_{\bar1}(q,{\vartheta}), q\in\Omega, {\vartheta}\in\rod^n\}.
$$
\par
(2) For a supersmooth function $u(x,\theta)=\sum_{|a|\le n}\theta^a u_a(x)$ defined on ${\mathfrak{M}}$,
we call the following expression  as ``the integral of the function $u(x,\theta)$ over the foliated singular manifold ${\mathfrak{M}}$";
\begin{equation}
\VViint_{\!\!\!\mathfrak{M}} dxd\theta\,u(x,\theta)
=\int_{\rod^n} d{\vartheta}\,\bigg[\int_{\Omega} dq\,\sdet J(\gamma)(q,{\vartheta})u(\gamma(q,{\vartheta}))\bigg].
\label{20100315}
\end{equation}
Here, we assume that for each $\eta\in\rod^n$, the integral in the bracket $[{\cdots}]$ above exists as the integral on $\Omega$.
\end{definition}

\begin{definition}
Let two foliated singular manifolds ${\mathfrak{M}}=\gamma(\widetilde{\Omega})$ and ${\mathfrak{M}}_1=\gamma_1(\widetilde{\Omega}_1)$ be given. We call these are superdiffeomorphic if
there exist a diffeomorphism $\phi:\widetilde{\Omega}_1\to \widetilde{\Omega}$ and $\varphi:{\mathfrak{M}}_1\to{\mathfrak{M}}$
such that $\gamma_1=\varphi^{-1}\circ \gamma\circ\phi$.
$$
{\begin{CD}
{\widetilde{\Omega}}@>\gamma>>{\mathfrak{M}}=\gamma({\widetilde{\Omega}})\\ 
@A{\phi}AA  
@AA{\varphi}A\\
{\widetilde{\Omega}}_1@>\gamma_1>>{\mathfrak{M}}_1=\gamma_1({\widetilde{\Omega}}_1).
\end{CD}}
$$
\end{definition}

\begin{theorem}\label{CVF-VV}
Let
\begin{equation}
\varphi:(y,\omega)\to (x,\theta)\with
x=\varphi_{\bar0}(y,\omega), \quad \theta=\varphi_{\bar1}(y,\omega)
\label{VV4.6}
\end{equation}
be a {supersmooth} diffeomorphism from
the neighbourhood ${\mathfrak{O}}_1$ of the foliated singular
manifold ${\mathfrak{N}}(\delta,\Omega)$ in $\supermn$ onto
the neighbourhood ${\mathfrak{O}}$ of the foliated singular
manifold ${\mathfrak{M}}(\gamma,\Omega)$ in $\supermn$,
that  is, ${\mathfrak{M}}=\varphi({\mathfrak{N}})$ and $\sdet J(\varphi)\neq0$.
We assume moreover that $\delta=\varphi^{-1}\circ\gamma$ with $\sdet J(\gamma)\neq0$.
\par
Then, for any function $u\in{\CSS}({\mathfrak{O}}:{\fR})$
which is integrable on ${\mathfrak{M}}$, we have CVF %the change of variables formula
\begin{equation}
\VViint_{\!\!\!\mathfrak{M}} dxd\theta\,u(x,\theta)=
\VViint_{\!\!\!\varphi^{-1}({\mathfrak{M}})} dyd\omega\,{\sdet}J(\varphi)(y,\omega){\cdot}u(\varphi(y,\omega)).
\label{VV4.10-1}
\end{equation}
\end{theorem}
\begin{remark} An analogous statement on Banach-Grassmann algebra ${\mathfrak{B}}_{L}^{m|n}$ is proved in \cite{VV84} under the condition that the set $\{x\in{\mathfrak{B}}_{L\bar0}^m\;|\; x=\gamma(q,{\vartheta}),\,q\in\Omega\}$ is independent  for each ${\vartheta}\in{\mathfrak{B}}_{L\bar1}^n$.
\end{remark}
\begin{remark} The formulas \eqref{B2.2.12} and \eqref{VV4.10-1} look same but their underlying definitions \eqref{BF} and \eqref{20100315} are different! This is related to the problem ``How to consider the body of supermanifolds?'' (see, Catenacci, Reina and Teofilatto~\cite{CRT}).
\end{remark}

\section{Illustration: resolution of inconsistency of Example \ref{Example1} by contour integral}
From Theorem \ref{CVF-VV}, we have the following interpretation:
\par
Let $\widetilde\Omega=\Omega\times\rod^2$ with $\Omega=(0,1)$ be given. 
Defining $\gamma:\widetilde{\Omega}\to{\mathfrak{M}}$ by
$$
\gamma:\widetilde\Omega\ni(q,{\vartheta})\to(x,\theta)=(\gamma_{\bar0}(q,{\vartheta}), \gamma_{\bar1}(q,{\vartheta}))=\gamma(q,{\vartheta}),
$$
we may consider ${\mathfrak{M}}=\{(x,\theta)\in{\fR}^{1|2}\;|\; \pi_{\mathrm{B}}(x)\in\Omega,\;\theta\in\rod^2\}$ as
a singular foliated manifold $\gamma(\widetilde{\Omega})$ in ${\fR}^{1|2}$.
Prepare another singular foliated manifold ${\mathfrak{N}}=\delta(\widetilde{\Omega})$ in ${\fR}^{1|2}$
with a superdiffeomorphism
$$
\varphi:\delta(\widetilde\Omega)\ni(y,\omega)\to\varphi(y,\omega)=(x,\theta)\in\gamma(\widetilde\Omega),
$$
given by
$$
\begin{cases}
x=\varphi_{\bar0}(y,\omega)=y+\omega_1\omega_2\phi(y),\\
\theta_1=\varphi_{\bar1,1}(y,\omega)=\omega_1,\, \theta_2=\varphi_{\bar1,2}(y,\omega)=\omega_2,
\end{cases}
$$ 
and
$$
\delta=\varphi^{-1}\circ\gamma:(q,{\vartheta})\to (q-{\vartheta}_1{\vartheta}_2\phi(q),{\vartheta})
=(\delta_{\bar0}(q,{\vartheta}),\delta_{\bar1}(q,{\vartheta}))=(y,\omega).
$$
Then, we have ${\mathfrak{N}}=\varphi^{-1}({\mathfrak{M}})$ and
$$
J(\varphi)(y,\omega)=\begin{pmatrix}
1+\omega_1\omega_2\phi'(y)&0&0\\
\omega_2\phi(y)&1&0\\
-\omega_1\phi(y)&0&1
\end{pmatrix},\;\;
J(\gamma)(q,{\vartheta})=
\begin{pmatrix}
1&0&0\\
0&1&0\\
0&0&1
\end{pmatrix},\;\;
J(\delta)(q,{\vartheta})=\begin{pmatrix}
1-{\vartheta}_1{\vartheta}_2\phi'(q)&0&0\\
-{\vartheta}_2\phi(q)&1&0\\
{\vartheta}_1\phi(q)&0&1
\end{pmatrix}.
$$
\par
In this case, for $u(x,\theta)=u_{\bar0}(x)+\theta_1\theta_2u_{\bar1}(x)$, we have
$$
\begin{aligned}
\VViint_{\!\!\!\mathfrak{M}} dxd\theta\, u(x,\theta)
&=\int_{\rod^2} d{\vartheta}\,\bigg[\int_\Omega dq \, {\sdet}J(\gamma)(q,{\vartheta})u(\gamma(q,{\vartheta}))\bigg]\\
&=\int_0^1dq\,\int_{\rod^2} d{\vartheta}\, u(q,{\vartheta})
=\int_0^1 dq\,
\frac{\partial}{\partial{\vartheta}_2}\frac{\partial}{\partial{\vartheta}_1}u(q,{\vartheta})
\bigg|_{{\vartheta}=0}=\int_0^1 dq\,u_{\bar1}(q),
\end{aligned}
$$
and
\begin{equation}
\begin{aligned}
\VViint_{\!\!\!{\mathfrak{N}}}dyd\omega\,(\varphi^*u)(y,\omega)
&=\iint_{\widetilde{\Omega}} dq d{\vartheta}\, \sdet J(\delta)(q,{\vartheta})
\big[\sdet J(\varphi)(y,\omega)u(\varphi(y,\omega))\big]_{(y,\omega)=\delta(q,{\vartheta})}\\
&=\int_0^1dq\,\bigg[\int_{\rod^2} d{\vartheta}\,\sdet J(\delta)(q,{\vartheta})
\big[\sdet J(\varphi)(y,\omega)u(\varphi(y,\omega))\big]_{(y,\omega)=\delta(q,{\vartheta})}\bigg]\\
&=\int_0^1dq\,\int_{\rod^2} d{\vartheta}\, u(q,{\vartheta}). 
\end{aligned}
\label{VV-C}
\end{equation}
Therefore, we have the following result with no condition on the support of $u$:
$$
\begin{aligned}
\VViint_{\!\!\!\mathfrak{M}} dxd\theta\, u(x,\theta)
&=\iint_{\widetilde{\Omega}}dq d{\vartheta}\,\sdet J(\gamma)(q,{\vartheta}){\cdot}u(\gamma(q,{\vartheta}))\\
&=\iint_{\widetilde{\Omega}}dq d{\vartheta}\,
\sdet J(\delta)(q,{\vartheta})\big[\sdet J(\varphi)(y,\omega){\cdot}u(\varphi(y,\omega))\big]\bigg|_{(y,\omega)=\delta(q,{\vartheta})}\\
&=\VViint_{\!\!\!\varphi^{-1}({\mathfrak{N}})}dyd\omega\,\sdet J(\varphi)(y,\omega){\cdot}u(\varphi(y,\omega))
=\VViint_{\!\!\!{\mathfrak{N}}}dyd\omega\,(\varphi^*u)(y,\omega).\qed
\end{aligned}
$$

\begin{remark}
For the future use, we calculate more precisely:
\begin{equation}
\begin{aligned}
\sdet J(\varphi)(y,\omega){\cdot}u(\varphi(y,\omega))
&=(1+\omega_1\omega_2\phi'(y))[u_{\bar0}(y+\omega_1\omega_2\phi(y))+\omega_1\omega_2u_{\bar1}(y+\omega_1\omega_2\phi(y))]\\
&=(1+\omega_1\omega_2\phi'(y))[u_{\bar0}(y)+\omega_1\omega_2(\phi(y)u_{\bar0}'(y)+u_{\bar1}(y))]\\
&=u_{\bar0}(y)+\omega_1\omega_2[\underline{(\phi(y)u_{\bar0}(y))'}+u_{\bar1}(y)],
\end{aligned}
\label{VV-in}
\end{equation}
and putting $(y,\omega)=\delta(q,{\vartheta})$, we have
\begin{equation}
\begin{aligned}
\sdet J(\delta)(q,{\vartheta})
&\big[\sdet J(\varphi)(y,\omega){\cdot}u(\varphi(y,\omega))\big]_{(y,\omega)=\delta(q,{\vartheta})}\\
&=(1-{\vartheta}_1{\vartheta}_2\phi'(q))\big(u_{\bar0}(y)+\omega_1\omega_2[(\phi(y)u_{\bar0}(y))'+u_{\bar1}(y)]\big)\big|_{\scriptstyle{y=q-{\vartheta}_1{\vartheta}_2\phi(q),}\atop\scriptstyle{\omega_1={\vartheta}_1,\,\omega_2={\vartheta}_2}}\\
&={\underline{(1-{\vartheta}_1{\vartheta}_2\phi'(q))}}\big[{\underline{u_{\bar0}(q)-{\vartheta}_1{\vartheta}_2\phi(q)u_{\bar0}'(q)}}+{\vartheta}_1{\vartheta}_2
[(\phi(q)u_{\bar0}(q))'+u_{\bar1}(q)]\big]\\
&=u_{\bar0}(q)+{\underline{{\vartheta}_1{\vartheta}_2}}[(\phi(q)u_{\bar0}(q))'+u_{\bar1}(q){\underline{-(\phi(q)u_{\bar0}(q))'}}]
=u_{\bar0}(q)+{\vartheta}_1{\vartheta}_2u_{\bar1}(q).
\end{aligned}
\end{equation}
Or  since $u(\varphi(y,\omega))_{(y,\omega)=\delta(q,{\vartheta})}=u(q,{\vartheta})$ and
$$
\sdet J(\delta)(q,{\vartheta}){\cdot}\sdet J(\varphi)(\delta(q,{\vartheta}))=(1-{\vartheta}_1{\vartheta}_2\phi'(q))(1+{\vartheta}_1{\vartheta}_2\phi'(q))=1,
$$
we have the result.
\par
Therefore, the appearance of the term $\omega_1\omega_2(\phi(y)u_{\bar0}(y))'$ in \eqref{VV-in}
is \underline{the very reason of inconsistency}.
\end{remark}

\section{Integration w.r.t. even variables}
\subsection{One dimensional case as a prototype}
We recall the idea of the contour integral noted in Rogers~\cite{rog86-2}.
\begin{quotation}
Contour integrals are a means of ``pulling back'' an integral in a space that is algebraically (as well as
possibly geometrically) more complicated than $\euc^m$. A familiar example, of course, is complex contour integration; 
if $\gamma:[0,1]\to{\mathbb{C}}$ is piecewise $C^1$ and $f:{\mathbb{C}}\to{\mathbb{C}}$, one has
the one-dimensional contour integral
$$
\int_{\gamma} f(z)dz=\int_0^1f(\gamma(t)){\cdot}\gamma'(t)dt=\int_0^1dt\,\gamma'(t){\cdot}f(\gamma(t)).
$$
This involves the algebraic structure of ${\mathbb{C}}$ because the right-hand side of above includes multiplication $\cdot$ of complex numbers.
\end{quotation}
We follow this idea to define the integral of a {supersmooth} function $u(x)$ 
on an even superdomain ${\mathfrak{U}}_{\mathrm{ev}} \subset \supermo=\rev^m$
 (see also, Rogers~\cite{rog85-1,rog85-2,rog86-3} and Vladimirov and Volovich~\cite{VV84}). 

\begin{definition}\label{4.1} 
Let $u(x)$ be a {supersmooth} function defined on an even superdomain 
${\mathfrak{U}}_{\mathrm{ev}}\subset{\fR}^{1|0}$ such that 
$[a,b]\subset \pi_{\mathrm{B}}({\mathfrak{U}}_{\mathrm{ev}})$.
Let $ \lambda=\lambda_{\mathrm B} + \lambda_{\mathrm S}$, 
$\mu =\mu_{\mathrm B} + \mu_{\mathrm S} \in {\mathfrak{U}}_{\mathrm{ev}}$ 
with $\lambda_{\mathrm B}=a$, $\mu_{\mathrm B}=b$,
and let a continuous and piecewise $C^1 $-curve 
$ \gamma : [a,b] \,\rightarrow \, {\mathfrak{U}}_{\mathrm{ev}} $
be given such that 
$\gamma(a) = \lambda $, $\gamma(b)  = \mu $.
We define
\begin{equation}
\int_{\gamma}  dx \, u(x)
=
\int_{a}^{b} dt\, \gamma'(t){\cdot}u(\gamma(t))\in{\fC}\with \gamma'(t)={\dot \gamma}(t)=\frac{d\gamma(t)}{dt}
\label{4-1}
\end{equation}
and call it the {integral of $u$} along the curve $\gamma$.
\end{definition} 

Using the integration by parts for functions on $\euc$, we get the following fundamental result. 
\begin{proposition} [p.7 of de Witt~\cite{DW2}]\label{4.2}
Let $u(t) \in C^{\infty} ([{a},{b}]:{\fC})$ and $U(t) \in C^{\infty} ([{a},{b}]:{\fC}) $ be given such that $ U' (t)=u(t) $ on
$ [ {a} , {b} ] $. We denote the Grassmann continuations of them  as  $\tilde{u}(x)$ and  $\tilde{U}(x)$. %$u(t)$ and $U(t)$
Then, for any continuous and piecewise
$C^1 $-curve ${\gamma}:[ {a} , {b} ] 
\, \rightarrow \, {\mathfrak{U}}_{\mathrm{ev}}\subset{\fR}^{1|0}$
such that $[ {a} , {b} ] \subset \pi_{\mathrm{B}}({\mathfrak{U}}_{\mathrm{ev}})$ and 
${\gamma}( {a} ) = \lambda$, ${\gamma}( {b} )=\mu $ with $\lambda_{\mathrm B}=a$, $\mu_{\mathrm B}=b$,
we have
\begin{equation}
\int_{\gamma}  dx \, \tilde{u}(x)= \tilde{U}( \lambda)-  \tilde{U}(\mu).
\label{4-2}
\end{equation}
\end{proposition}
\par
{\it Proof}.
Denoting ${\dot{\gamma}_{\mathrm B}}(t)=\dt {\gamma}_{\mathrm B}(t)$, etc.  and by definition, we get
$$
\begin{aligned}
\int_{{a}}^{{b}}dt\,{\gamma}'(t)u({\gamma}(t))
&=
\int_{{a}}^{{b}}dt\,({\dot{\gamma}_{\mathrm B}}(t)+{\dot{\gamma}_{\mathrm S}}(t))
\sum_{\ell\ge 0} {\frac 1{\ell!}}
u^{(\ell)}({\gamma}_{\mathrm B}(t)){\gamma}_{\mathrm S}(t)^\ell\\
{}
&=\int_{{a}}^{{b}}dt\,
{\dot{\gamma}_{\mathrm B}}(t)u({\gamma}_{\mathrm B}(t))
+\int_{{a}}^{{b}}dt\,
{\dot{\gamma}_{\mathrm B}}(t)\sum_{k\ge 1} {\frac 1{k!}}
u^{(k)}({\gamma}_{\mathrm B}(t)){\gamma}_{\mathrm S}(t)^k\\
&\qquad\qquad\qquad\qquad\qquad\qquad\qquad\qquad
+\int_{{a}}^{{b}}dt\,
\sum_{\ell\ge 0} {\frac 1{\ell!}}
u^{(\ell)}({\gamma}_{\mathrm B}(t)){\dot{\gamma}_{\mathrm S}}(t){\gamma}_{\mathrm S}(t)^\ell\\
&=
U({b})-U({a})+
\sum_{\ell\ge 0} {\frac 1{(\ell+1)!}}
\left\{
U^{(\ell+1)}({b})\mu_{\mathrm S}^{\ell+1}
-U^{(\ell+1)}({a})\lambda_{\mathrm S}^{\ell+1}
\right\}\\
&=\tilde{U}(\mu)-\tilde{U}(\lambda).	
\end{aligned}
$$
Here, we used the integration by parts formula for functions on $\euc$ valued in Fr\'echet space~\cite{ham82}:
$$
\begin{aligned}
\int_{{a}}^{{b}}dt\,
&u^{(\ell)}({\gamma}_{\mathrm B}(t)){\dot{\gamma}_{\mathrm S}}(t){\gamma}_{\mathrm S}(t)^\ell\\
&=
\int_{{a}}^{{b}}dt\,
u^{(\ell)}({\gamma}_{\mathrm B}(t))
\dt \frac{{\gamma}_{\mathrm S}(t)^{\ell+1}}{\ell+1}\\
&=-\int_{{a}}^{{b}}dt\,
{\dot{\gamma}_{\mathrm B}}(t)u^{(\ell+1)}({\gamma}_{\mathrm B}(t))
\frac{{\gamma}_{\mathrm S}(t)^{\ell+1}}{\ell+1}
+U^{(\ell+1)}({{b}})\frac{{\mu_{\mathrm S}}^{\ell+1}}{\ell+1}
-U^{(\ell+1)}({{a}})\frac{{\lambda_{\mathrm S}}^{\ell+1}}{\ell+1}.  \qed
\end{aligned}
$$

\begin{remark}
Unless there occurs the confusion, we denote simply $\tilde{u}(x)$,  $\tilde{U}(x)$ as $u(x)$, $U(x)$, respectively.
\end{remark}

\begin{lemma}[Lemma 3.9 in \cite{rog85-1} on ${\mathfrak{B}}_L$]\label{rog3.9}
(a) (reparametrization of paths) Let $\gamma:[a,b]\to\rev$ be a path in $\rev$ and let $c,\,d\in\euc$.
Also let $\phi:[c,d]\to[a,b]$ be $C^1$ with $\phi(c)=a$, $\phi(d)=b$ and $\phi'(s)>0$ for all $s\in[c,d]$.
Then
$$
\int_{\gamma} dx\, u(x)=\int_{\gamma\circ\phi}dx\, u(x).
$$
(b)(sum of paths) Let $\gamma_1:[a,b]\to\rev$ and $\gamma_2:[c,d]\to\rev$ be two paths with $\gamma_1(b)=\gamma_2(c)$ Also define $\gamma_1+\gamma_2$ to be the path $\gamma_1+\gamma_2:[a,b+d-c]\to \rev$ defined by
$$
\gamma_1+\gamma_2(t)=\begin{cases}
\gamma_1(t), &{a\le t\le b},\\
\gamma_2(t-b+c), & {b\le t\le b+d-c}.
\end{cases}
$$
Then if ${\mathfrak{U}}_{\mathrm{ev}}$ is open in $\rev$, $u:{\mathfrak{U}}_{\mathrm{ev}}\to{\fR}$ is in $\CSS$ and $\gamma_1([a,b])\subset {\mathfrak{U}}_{\mathrm{ev}}$, $\gamma_2([c,d])\subset {\mathfrak{U}}_{\mathrm{ev}}$,
$$
\int_{\gamma_1+\gamma_2}dx\,u(x) =\int_{\gamma_1}dx\,u(x)+\int_{\gamma_2}dx\,u(x).
$$
(c)(inverse of a path)  Let $\gamma:[a,b]\to\rev$ be a path in $\rev$.
Define the curve
$-\gamma:[a,b]\to\rev$ by
$$
-\gamma(t)=\gamma(a+b-t)
$$
Then if ${\mathfrak{U}}_{\mathrm{ev}}$ is open in $\rev$
 with $\gamma([a,b])\subset {\mathfrak{U}}_{\mathrm{ev}}$ and $u:{\mathfrak{U}}_{\mathrm{ev}}\to\fR$ is supersmooth,
 $$
 \int_{-\gamma} dx\, u(x)= -\int_{\gamma} dx\,u(x).
 $$
 \end{lemma}
 \par {\it Proof}.
Applying CVF on $\euc$ for $t=\phi(s)$ and $dt=\phi'(s)ds$, we have
 $$
 \begin{aligned}
\int_{\gamma\circ\phi}dx\, u(x)=\int_c^d ds\, (\gamma(\phi(s)))'u(\gamma(\phi(s)))
&=\int_c^d ds\, \phi'(s)[\gamma'(\phi(s))u(\gamma(\phi(s)))]\\
&=\int_a^b dt\,\gamma'(t)u(\gamma(t))=\int_{\gamma} dx\, u(x).
\end{aligned}
$$
Other statements are also proved analogously. $\qquad\square$ %\qed

\begin{corollary}[Corollary 3.7 in \cite{rog85-1} on ${\mathfrak{B}}_L$]\label{4.3}
Let $u(x)$ be a {supersmooth} function defined on a even superdomain 
${\mathfrak{U}}_{\mathrm{ev}} \subset {\fR}^{1|0}$ into $\fC$.\\
(a) Let $ {\gamma}_1 , {\gamma}_2 $ be continuous and piecewise $C^1 $-curves from 
$[a,b] \, \rightarrow \, {\mathfrak{U}}_{\mathrm{ev}}$
such that $\lambda ={\gamma}_1(a)={\gamma}_2 (a) $ 
and $ \mu = {\gamma}_1 (b) ={\gamma}_2 (b) $.
If ${\gamma}_1 $ is homotopic to ${\gamma}_2 $, then 
\begin{equation}
\int_{{\gamma}_1}dx \, u(x)
= \int_{{\gamma}_2}dx \, u(x).
\label{4-3}
\end{equation}
(b) If $u:\rev\to{\fR}$ is $\CSS$ on all $\rev$, one denote it ``unambiguously'' as
$$
\int_{\lambda}^{\mu}dx\, u(x)=\int_{\gamma} dx\,u(x).
$$
Here, $\gamma:[a,b]\to \rev$ is any path in $\rev$ with $\gamma(a)={\lambda}$, $\gamma(b)={\mu}$.
\end{corollary}

\begin{proposition}
For a given change of variable $x=\varphi(y)$, 
we define the pull-back  of 1-form ${\mathfrak{v}}_x=dx\,\rho(x)$ 
by
$(\varphi^*{\mathfrak{v}})_y=dy\,\frac{\partial \varphi(y)}{\partial y}\,\rho(\varphi(y))$. 
Then, for paths $\gamma:[a,b]\to \fR^{1|0}_x$, $\varphi^{-1}\circ\gamma:[a,b]\to \fR^{1|0}_y$ and
$u$, we have
$$
\int_{\gamma}\,{\mathfrak{v}}=\int_{\gamma}dx\,{\mathfrak{v}}_x\,\rho(x)u(x)
=\int_{\varphi^{-1}\circ\gamma}dy\,(\varphi^*{\mathfrak{v}})_y\,\rho(\varphi(y))u(\varphi(y))
=\int_{\varphi^{-1}\circ\gamma}\varphi^*{\mathfrak{v}}\,\varphi^*u.
$$
\end{proposition}
\par
{\it Proof}. From
\begin{equation}
{\begin{CD}
[a,b]@>\gamma>>{{\fR}_{x}^{1|0}}\\ %\overset{u}{\to} {\fR}\\ %=\gamma(\widetilde{\Omega})\\
@|@AA{\varphi}A\\
[a,b]@>\delta>>{{\fR}_{y}^{1|0}}
\end{CD}}
\quad\with
\delta=\psi\circ\gamma\et \psi={\varphi}^{-1},
\label{R3.8}
\end{equation}
we have not only
$$
\int_{\gamma} {\mathfrak{v}}_x u(x)=\int_a^b dt\, \dot{\gamma}(t)\rho(\gamma(t))u(\gamma(t)),
$$
but also
$$
\begin{aligned}
\int_{\varphi^{-1}\circ\gamma}(\varphi^*{\mathfrak{v}})_y\varphi^*u(y)
&=\int_{\varphi^{-1}\circ\gamma} dy\, \frac{d\varphi(y)}{d y} \rho(\varphi(y))u(\varphi(y))\\
&=\int_a^b dt\, \dt(\varphi^{-1}(\gamma(t)))\frac{d\varphi(y)}{d y}\rho(\varphi(y)) u(\varphi(y))\bigg|_{y=\varphi^{-1}\circ\gamma(t)}
=\int_a^b dt\, \gamma'(t)\rho(\gamma(t))u(\gamma(t)).
\end{aligned}
$$
Here, we used $y=\varphi^{-1}(\varphi(y))=\psi(\varphi(y))$, $x=\gamma(t)$, $y=\psi(\gamma(t))$ with
$$
1=\frac{d\varphi(y)}{dy}{\cdot}\frac{d\psi(x)}{dx}\bigg|_{x=\varphi(y)},\;
 \dt(\psi(\gamma(t)))={\gamma'}(t)\frac{d\psi(x)}{dx}\bigg|_{x=\gamma(t)},
\quad
\frac{d\psi(x)}{dx}\bigg|_{x=\varphi(y)}=
\bigg(\frac{d\varphi(y)}{d y}\bigg)^{-1}. \qed
$$

\begin{example}[Translational invariance]\label{TI}
Let $I=(a,b)\subset \euc$.  
We put
${\mathfrak{M}}=\gamma(I)=\{x\in\rev\;|\; \pi_{\mathrm{B}}(x)=q\in I\}\subset\rev$
 by identifying $q\in I$ as $\gamma(q)=x\in\rev$.%={\mathfrak{M}}(\iota, I)
Taking a non-zero nilpotent element $\nu\in\rev$, that is, $0\neq\nu$ and $\pi_{\mathrm{B}}(\nu)=0$, we put $\tau_{\nu}:\rev\ni y\to x=\varphi(y)=\tau_{\nu}(y)=y-\nu\in\rev$,
$$
{\mathfrak{M}}_1=\tau_{\nu}^{-1}({\mathfrak{M}})=\{x+\nu\rev\;|\; \pi_{\mathrm{B}}(x)=q\in I\},\quad \gamma_1(q)=\tau_{\nu}^{-1}(\gamma(q)).
$$
Then, we have
$$
\int_{{\mathfrak{M}}}dx\,u(x)=\int_a^b dq\,\gamma'(q)u(\gamma(q))=\int_a^b dq \,\gamma_1'(q)u(\gamma(q))
=\int_{{\mathfrak{M}}_1}dy\,u(y-\nu).
$$
\end{example}
\begin{remark} %
(i) Above identification $\gamma(q)=x\in\rev$ is obtained as the Grassmann continuation $\tilde{\iota}$ of 
a function $\iota(q)=q\in C^{\infty}(I:\euc)$. In fact,
$$
\tilde{\iota}(x)=\sum_{\alpha}\frac{\partial^{\alpha}\iota(q)}{\partial q^{\alpha}}(x_{\mathrm{B}})x_{\mathrm{S}}^{\alpha}
=x_{\mathrm{B}}+x_{\mathrm{S}}=x.
$$
(ii) As is noted in Example 2.2 of  ~\cite{rog85-1}, there occurs an inconsistency when we apply the naive definition of integration \eqref{BF}: Let ${\mathfrak{U}}_{\mathrm{ev}}=\pi_{\mathrm{B}}^{-1}(a,b)$.
Then we have
$$
\int_{\mathfrak{U}_{\mathrm{ev}}}dx\,x=\int_a^b dq\,q=\frac{1}{2}(b^2-a^2).
$$
For $0\neq\nu\in\rev$ with $\pi_{\mathrm{B}}(\nu)=0$, we have $\{x=y-\nu\;|\; y\in\pi_{\mathrm{B}}^{-1}(a,b)\}={\mathfrak{U}}_{\mathrm{ev}}$,  $``dx=dy''$ and therefore
$$
\int_{\mathfrak{U}_{\mathrm{ev}}}dy\, (y-\nu)=\int_a^b\, dq (q-\nu)=\frac{1}{2}(b^2-a^2)-\nu(b-a)\neq\int_{\mathfrak{U}_{\mathrm{ev}}}dx\,x.
$$
This inconsistency stems from the naive definition \eqref{BF}, that is, Berezin's integral w.r.t. even variables is an integral over $\euc^m$ and \underline{not} on $\rev^m$. 
By the nilpotency of $\nu$, we have
$$
(\nu+\tilde{b})^2-(\nu+\tilde{a})^2=\tilde{b}^2-\tilde{a}^2+2\nu(\tilde{b}-\tilde{a})=b^2-a^2+2\nu(b-a),
$$
which remedies this inconsistency by formally putting
$$
\int_{{\mathfrak{M}}_1}\,dy\, u(\varphi(y))=\int_{\tilde{a}+\nu}^{\tilde{b}+\nu}\,dy \, u(\varphi(y))=\frac{1}{2}(b^2-a^2).
$$
Or more rigorous description is given in Example \ref{TI} above.
\end{remark}

\subsection{Many dimensional case} %Vladimirov and Volovich's approach: 
\subsubsection{\`a la Rogers~\cite{rog85-1}}
We replace ${\mathfrak{B}}_{L}$ or ${\mathfrak{B}}$ with ${\fR}$ in her arguments.
\begin{definition}[$m$-path]
Let ${\bf{I}}^m=\prod_{j=1}^m[a_j,b_j]\subset\euc^m$. For a continuous and piecewise $C^1$ function $\gamma(t):{\bf{I}}^m\to {\fR}_{\mathrm{ev}}^m$ (called $m$-path) with
$\gamma(t)=(\gamma_1(t),{\cdots},\gamma_m(t))$, $t=(t_1,{\cdots},t_m)$,
we define
$$
\int_{\gamma}dx_1{\cdots}dx_m\, u(x)=\int_{{\bf{I}}^m}dt_1{\cdots}dt_m \det(J(\gamma)(t))u(\gamma(t)).
$$
Here, $u:{\mathfrak{U}}_{\mathrm{ev}}\to{\fR}$ is continuous on an open set ${\mathfrak{U}}_{\mathrm{ev}}\subset{\fR}_{\mathrm{ev}}^m$ containing $\gamma({\bf{I}}^m)$ and
$$
J(\gamma)(t)=\bigg(\frac{\partial\gamma_i(t)}{\partial t_j}\bigg)_{i,j=1,{\cdots},m}.
$$
\end{definition}

\begin{proposition}[see, Theorem 4.4 of Rogers~\cite{rog85-1} on ${\mathfrak{B}}_L$]
Let ${\mathfrak{U}}_{\mathrm{ev}}$ be open in ${\fR}_{\mathrm{ev}}^m$ and let ${\phi}:{\mathfrak{U}}_{\mathrm{ev}}\to{\fR}_{\mathrm{ev}}^m$ be an injective and supersmooth mapping.
And let $\gamma:{\bf{I}}^m\to {\mathfrak{U}}_{\mathrm{ev}}$ be an $m$-path in ${\mathfrak{U}}_{\mathrm{ev}}$ and $u:{\mathfrak{U}}_{\mathrm{ev}}\to{\fR}$ be continuous. Then
$$
\int_{\gamma} dx u=\int_{\phi\circ\gamma} dx\, (\det J(\phi))^{-1}{\cdot} u\circ  \phi^{-1}
\with
J(\phi)=\frac{\partial {\phi}_i}{\partial x_j})
$$
with
${\phi}(x)=({\phi}_1(x),{\cdots}, {\phi}_m(x))$, $x=(x_1,{\cdots}, x_m)$.
\end{proposition}

%% input p-forms@@

\subsubsection{\`a la Vladimirov and Volovich~\cite{VV84}} Since they use ${\mathfrak{B}}_{L}$ in their arguments, we need some modifications
to work on $\fR$.
\begin{definition}[A $m$-dimensional singular manifold and integrals on it]
Let ${\mathfrak{M}}$ be $m$-dimensional singular manifold, that is, there exists a pair $({M},\gamma)$ such that ${\mathfrak{M}}=\gamma({M})$ where ${M}$ is an oriented region in $\eucm$ 
and $\gamma:{M}\to\gamma({M})\subset\rev^m$.
For a given %${\mathfrak{v}}\in\Omega_m^{(0)}({\mathfrak{M}}:\fR)$, i.e.
${\mathfrak{v}}=dx_1\wedge{\cdots}\wedge dx_m\,\rho(x)$ with
$\rho\in{\CSS}({\mathfrak{M}}:\fR)$
and $u\in \CSS({\mathfrak{M}}:\fR)$, if the integral of the right-hand side exists, we define
\begin{equation}
\int_{\mathfrak{M}}{\mathfrak{v}}u
=\int_{\mathfrak{M}} {\mathfrak{v}}_x\,u(x)
=\int_M \gamma^*{\mathfrak{v}}{\cdot} \gamma^*u
=\int_{M} dq\,\det\frac{\partial\gamma(q)}{\partial q}\, \rho(\gamma(q)){\cdot}u(\gamma(q)),
\label{090222}
\end{equation}
and if $u=1$, then the form ${\mathfrak{v}}$ is said to be integrable over the singular manifold ${\mathfrak{M}}=\gamma(M)$.
\end{definition}

\begin{definition}
The pairs  $(M,\gamma)$ and  $(N,\delta)$ are said to be
equivalent if ${\mathfrak{M}}=\gamma(M)=\delta(N)$ and there exists a diffeomorphism
$\phi:N\to M$ such that $\delta=\gamma\circ \phi$. Thus
\begin{equation}
{\begin{CD}
M@>\gamma>>{\mathfrak{M}}\\ %=\gamma(\widetilde{\Omega})\\
@A{\phi}AA  %
@|\\
%@VV{\cong}V\\
N@>\delta>>{\mathfrak{N}}
\end{CD}}\quad
\Longrightarrow
\int_M\gamma^*{\mathfrak{v}}=\int_{N}\delta^*{\mathfrak{v}}.
\label{VVII2.2}
\end{equation}
\end{definition}
This implies that not only  the integral \eqref{090222} doesn't depend on the choice of the pair $(M,\gamma)$ in an equivalent class but also we may interprete the formula \eqref{VVII2.2} 
as a change of variables formula as follows:

Let $\varphi$ be a mapping of class $C^1({\mathfrak{O}})$ of the neighborhood ${\mathfrak{O}}$ of ${\mathfrak{N}}$ in $\rev^m$ and $\varphi^*{\mathfrak{v}}$ is the pull-back of the superform ${\mathfrak{v}}$ under the mapping $\varphi$. Then
\begin{equation}
{\begin{CD}
M@>\gamma>>({\mathfrak{M}},{\mathfrak{v}})\\ 
@A{\phi}AA  %
@AA{\varphi}A\\
N@>\delta>>({\mathfrak{N}},\varphi^*{\mathfrak{v}})
\end{CD}}\;
\Longrightarrow
{\begin{aligned}
\int_{\mathfrak{M}}{\mathfrak{v}}&=\int_M dq\det J(\gamma)(q){\cdot}(\gamma^*v)_1\\
&=\int_N dq' \det J(\gamma\circ\phi)(q'){\cdot}(\gamma\circ\phi)^*{\mathfrak{v}}=\int_{{\mathfrak{N}}}\varphi^*{\mathfrak{v}}.
\end{aligned}}
\label{VVII2.3}
\end{equation}
That is, we have ${\phi^{-1}}^*\delta^*\varphi^*=\gamma^*$.

Restricting above argument to the case when $\gamma=\delta=\tilde{\iota}$ and $\varphi=\mbox{Id}$, the equality \eqref{VVII2.3} reduces to the ordinary change of variables formula:
$$
\int_M dq\,\rho(q)u(q)=\int_{N}dq'\,\det J(\phi)(q'){\cdot}\rho(\phi(q'))u(\phi(q'))\with {\mathfrak{v}}_x=dx\,\rho(x).
$$

\begin{example}[One-dimensional singular manifold]
Let $M=(0,1)$.
The integral of the superform
$$
{\mathfrak{v}}=\sum_{i=1}^m dx_i {\rho}_i(x)
$$
of degree $1$ along the curve ${\mathfrak{M}}=\{x=\gamma(q)\;|\; q\in M\}\subset\rev^m$ where $\gamma\in C^1((0,1):\rev^m)$, is determined by
$$
\int_{\mathfrak{M}} {\mathfrak{v}}=\sum_{i=1}^m\int_{M}dq\, \varphi'(q){\rho}_i(\varphi(q))
=\int_M \varphi^*{\mathfrak{v}}.
$$ 
\end{example}

\paragraph{\bf{A composition of change of variables}}
Let ${\mathfrak{V}}$ be a domain in $\rev^m$ and $\varphi^{(1)}:\mathfrak{V}\to\tilde{\mathfrak{U}}$ be a diffeomorphism. Moreover,
${\mathfrak{U}}$ be a domain in $\rev^m$ and $\varphi^{(2)}:\tilde{\mathfrak{U}}\to {\mathfrak{U}}$ be a diffeomorphism. 
$$
{\begin{CD}
U@>\gamma>>({\mathfrak{U}}, {\mathfrak{v}}),\\
%\;{\mathfrak{v}}_x=dx\,\rho(x),\\ %=\gamma(\widetilde{\Omega})\\
@A{\phi^{(2)}}AA  %
@AA{\varphi^{(2)}}A\\
%@VV{\cong}V\\
\tilde{U}@>\tilde\gamma>>(\tilde{\mathfrak{U}}, \tilde{\mathfrak{v}}),\\
%\; \tilde{\mathfrak{v}}_{\tilde{y}}=(\varphi^{{(2)}\,*}{\mathfrak{v}})_{\tilde{y}}=d{\tilde{y}}\,\det J(\varphi^{(1)})({\tilde{y}})\,\rho(\varphi^{(2)}({\tilde{y}}))),\\
@A{\phi^{(1)}}AA  %
@AA{\varphi^{(1)}}A\\
V@>\delta>>({\mathfrak{V}}, {\mathfrak{w}}),\\
%\;  {\mathfrak{w}}_y=dy\,(\varphi^{{(2)}\,*}\varphi^{{(2)}\,*}{\mathfrak{v}})_{\tilde{y}}
%=dy\,\det J(\varphi)(y)\rho(\varphi(y)).
\end{CD}}
\with
{\begin{aligned}
&{\mathfrak{v}}_x=dx\,\rho(x),\\ 
&\tilde{\mathfrak{v}}_{\tilde{y}}=(\varphi^{{(2)}\,*}{\mathfrak{v}})_{\tilde{y}}=d{\tilde{y}}\,\det J(\varphi^{(2)})({\tilde{y}}){\cdot}\rho(\varphi^{(2)}({\tilde{y}}))),\\
 &{\mathfrak{w}}_y=dy\,(\varphi^{{(1)}\,*}\varphi^{{(2)}\,*}{\mathfrak{v}})_{\tilde{y}}
=dy\,\det J(\varphi)(y){\cdot}\rho(\varphi(y)),\\
&\qquad\qquad\where \varphi=\varphi^{(2)}{\circ}\varphi^{(1)}.
\end{aligned}}
$$
Then we have,
$$
\int_{\tilde{\mathfrak{U}}}d\tilde{y}\, \rho(\tilde{y})v(\tilde{y})=\int_{{\mathfrak{V}}} dy\,\det J(\varphi^{(1)})(y){\cdot}v(\varphi^{(1)}(y))
\for v\in\CSS(\tilde{\mathfrak{U}}:{\fR})
$$
and
$$
\int_{\mathfrak{U}} dx\,u(x)=\int_{\tilde{\mathfrak{U}}}\,\det J(\varphi^{(2)})(\tilde{y}){\cdot}u(\varphi^{(2)}(\tilde{y}))
\for u\in\CSS({\mathfrak{U}}:{\fR}).
$$
Since $\det J(\varphi^{(1)})(y){\cdot}\det J(\varphi^{(2)})(\tilde{y})\big|_{\tilde{y}=\varphi^{(1)}(y)}=\det J(\varphi^{(2)}{\circ}\varphi^{(1)})(y)$, we have
$$
\int_{\mathfrak{U}} dx\,u(x)=\int_{\mathfrak{V}} dy\,\det J(\varphi^{(2)}{\circ}\varphi^{(1)})(y){\cdot}u(\varphi^{(2)}{\circ}\varphi^{(1)}(y)).
$$
Therefore, we have
$$
\int_{\varphi({\mathfrak{V}})} dx\,u(x)=\int_{\mathfrak{V}} dy\,\det J(\varphi)(y){\cdot}u(\varphi(y)).
$$

%%%%%
\section{Integration w.r.t. odd variables}
It seems natural to put formally
$$
d\theta_j=\sum_{{\bf{I}}\in{\bf{I}},|{\bf{I}}|={\mathrm{od}}}d\theta_{j,{\bf{I}}}\,\sigma^{\bf{I}}
\for \theta_j=\sum_{{\bf{J}}\in{\bf{J}},|{\bf{J}}|={\mathrm{od}}}\theta_{j,{\bf{J}}}\,\sigma^{\bf{J}}.
$$
\begin{remark}
Since above sum $\sum_{\bf{I}}$ stands for the position in the sequence space $\omega$ of K\"othe and the element of it is given by $d\theta_{j,{\bf{J}}}$ for $|{\bf{J}}|$ is finite, we may give the meaning to $d\theta_j$.
\end{remark}
Then, we have
$$
d\theta_j\wedge d\theta_k=d\theta_k\wedge d\theta_j.
$$
This makes us imagine that even if there exists the notion of integration,
it differs much from the standard one on $\euc^m$.

\subsection{Berezin integral} 
We follow Vladimirov and Volovich~\cite{VV84}, modifying it if necessary.
Since the supersmooth functions on ${\fR}^{0|n}$ are characterized as the polynomials with value in ${\fC}$, 
we need to define the integrability for those under the conditions that
\par
(i) integrability of all polynomials,  
\par 
(ii) linearity of an integral, and  
\par
(iii) invariance of the integral w.r.t. shifts.\\
Put ${\mathcal{P}}_n={\mathcal{P}}_n({\fC})=\{u(\theta)=\sum_{a\in\{0,1\}^n}\theta^au_a\;|\; u_a\in{\fC}\}$.

We say a mapping $I_{n}:{\mathcal{P}}_n\to{\fC}$ is an integral if it satisfies
%\par
%(1) ${\fR}$-linearity (from the right): $I(u\alpha+v\beta)=I(u)\alpha +I(v)\beta $ for $\alpha, \beta\in{\fR}$, $u,v\in{\mathcal{P}}_n$.
\par
(1) ${\fC}$-linearity (from the right): $I_{n}(u\alpha +v\beta )=I_{n}(u)\alpha+I_{n}(v)\beta$ for $\alpha, \beta\in{\fC}$, $u,v\in{\mathcal{P}}_n$.
\par
(2) translational invariance: $I_{n}(u({\cdot}+\omega))=I_{n}(u)$  for all $\omega\in {\fR}^{0|n}$ and $u\in{\mathcal{P}}_n$.

\begin{theorem}
For the existence of the integral $I_{n}$ satisfying above conditions (1) and (2), it is necessary and sufficient that
\begin{equation}
I_{n}(\phi_a)=0 \for \phi_a(\theta)=\theta^a,\;|a|\le n-1.
\label{VVII3.4}
\end{equation}
Moreover, we have
$$
I_{n}(u)=\frac{\partial}{\partial \theta_n}{\cdots}\frac{\partial}{\partial \theta_1}u(\theta)\bigg|_{\theta=0}I_{n}(\phi_{\bar{1}})
\where \phi_{\bar{1}}(\theta)=\theta^{\bar{1}}=\theta_1{\cdots}\theta_n.
$$
\end{theorem}
\par
{\it Proof}.
If there exists $I_{n}$ satisfying (1) and (2), then we have
$$
I_{n}(v)=\sum_{|a|\le n}I_{n}(\phi_a) v_a\for v(\theta)=\sum_{|a|\le n}\theta^a v_a=\sum_{|a|\le n}\phi_a(\theta) v_a.
$$
As
$$
\begin{aligned}
&(\theta+\omega)^a=\theta^a+\sum_{|a-b|\ge 1,b\le a}(-1)^*\theta^b\omega^{a-b},\\
&I_{n}(v({\cdot}+\omega))=\sum_{|a|\le n}I_{n}(\phi_a({\cdot}+\omega))v_a=\sum_{|a|\le n}I_{n}(\phi_a)v_a
+\sum_{|a|\le n}\sum_{|a-b|\ge 1,b\le a}(-1)^*I_{n}(\phi_b)v_b\omega^{a-b},
\end{aligned}
$$
by virtue of (2), we have
$$
\sum_{|a|\le n}\sum_{|a-b|\ge 1,b\le a}(-1)^*I_{n}(\phi_b)v_b\omega^{a-b}=0.
$$
Here, $v_a\in{\fC}$ and $\omega\in\rod^n$ are arbitrary, we have \eqref{VVII3.4}.
Converse is obvious. \qed

\begin{definition}\label{4.5}
We put $I_{n}(\phi_{\bar{1}})=1$, i.e.,
$$
\int_{\superon} \!\!\!\!d \theta_n \cdots  d \theta_1 
\,\theta_1 \cdots \theta_n =1.
\label{4-8}
$$
Therefore,  we put, for any $ v=\sum_{|a|{\le} n}\theta^a v_a \in {\mathcal{P}}_n (\fC) $
$$ 
I_{n}(v)=\int_{\superon} d \theta \, v( \theta ) 
= \int_{{\superon}}  
d \theta_n \cdots d \theta_1 \, v(\theta_1,\cdots ,\theta_n ) 
= (\partial_{\theta_n} \cdots 
\partial_{\theta_1} v)(0)=v_{\bar{1}}=\int_{\mathrm{Berezin}} d^n\theta f(\theta).
\label{4-7}
$$
This is called the (Berezin) integral of $v$ on  $ \superon$.
\end{definition}
Then, we have
\begin{proposition}\label{4.6}
Given
$ v, \, w \in {\mathcal{P}}_n ({\fC}) $ , we have the following:
\par
(1) (${\fC}$-linearity) 
For any homogeneous $ \lambda ,\, \mu \in {\fC}$,
\begin{equation}
\int_{\superon}  d \theta( \lambda v + \mu w)( \theta ) =
(-1)^{np(\lambda)}\lambda \int_{\superon} 
 d\theta \,v( \theta ) +
(-1)^{np(\mu)}\mu \int_{\superon} d \theta \,w( \theta ) .
\label{4-9}
\end{equation}
\par
(2) (Translational invariance) For any $\rho\in\superon$, we have
\begin{equation}
\int_{\superon}  d \theta \,v( \theta + \rho )=
\int_{\superon}  d \theta \,v( \theta ).
\label{4-10}
\end{equation}
\par
(3) (Integration by parts)
For $ v \in {\mathcal{P}}_n ( {\fC})$ such that $p(v) = 1$ or $0$, we have
\begin{equation}
\int_{\superon}\!\!\!\!d\theta 
\,v( \theta ) \partial_{\theta_s} w( \theta ) =
-(-1)^{p(v)}
\int_{\superon}\!\!\!\!d\theta \,(\partial_{\theta_s} v( \theta )) w( \theta ).
\label{4-11}
\end{equation}
\par
(4) (Linear change of variables)
Let $A=(A_{jk})$ with $A_{jk}\in \rev$ be an invertible matrix.
Then, 
\begin{equation}
\int_{\superon} \!\!\!\!d \theta \,v( \theta )
= (\det A )^{-1} \int_{\superon} \!\!\!\!d\omega \,v(A\cdot\omega ).
\label{4-12}
\end{equation}
\par
(5) (Iteration of integrals)
\begin{equation}
\int_{\superon}  d \theta \,v( \theta )
= \int_{{\mathfrak R}{}^{0|n-k}} \!\!\!\!d \theta_n  \cdots  d \theta_{k+1}
 \left( \int_{{\mathfrak R}{}^{0|k}} \!\!\!\!d \theta_{k}  \cdots d \theta_1
\,v(\theta_1, \cdots , \theta_k , \theta_{k+1} , \cdots , \theta_n ) \right).
\label{4-13}
\end{equation} 
\par
(6) (Odd change of variables)
Let $\theta=\theta(\omega)$ be an odd change of variables such that
$\theta(0)=0$ and
$\det\left.
\displaystyle{{\frac{\partial\theta(\omega)}{\partial \omega}}}
\right|_{\omega=0}\ne 0$.
Then, for any  $ v \in {\mathcal{P}}_n ({\fC})$,
\begin{equation}
\int_{\superon}  d \theta \, v(\theta)=
\int_{\superon}  d \omega 
\,\bigg({\det}{\frac{\partial \theta(\omega)}{\partial \omega}}\bigg)^{-1} v(\theta(\omega)).
\label{4-14}
\end{equation} 
\par
(7) ($\delta$-function) For $ v \in {\mathcal{P}}_n ({\fC})$ and $ \omega \in \superon $,
\begin{equation}
\int_{\superon}  d \theta 
\,( \omega_1 - \theta_1 ){\cdots}( \omega_n -  \theta_n ) v( \theta ) 
= v( \omega ) .
\label{4-15}
\end{equation} 
\eqref{4-15} allows us to put 
$$ \!\!\!
\delta ( \theta - \omega ) = ( \theta_1 - \omega_1 ) \cdots ( \theta_n - \omega_n ),
$$
though $\delta (- \theta )=(-1)^n \delta (\theta)$. 
\end{proposition}
\par
We omit the proof, since we may apply  the arguments
in pp.755-757 of Vladimirov and Volovich~\cite{VV84}
with slight modifications if necessary.

\begin{remark}
(i) We get the integration by parts formula, without the fundamental theorem of elementary analysis.\\
(ii) Moreover, since in conventional integration we get $\displaystyle{\int dy f(y)}=a\int dx f(ax)$, therefore the formula in \eqref{4-12} is very different from usual one. Analogous difference appears in \eqref{4-14}.
\end{remark}

%%%%%
\section{Integration w.r.t. even and odd variables}
\subsection{Proof of Theorem \ref{B1}}
For future use, we give a precise proof of Berezin~\cite{bere87} and Rogers~\cite{rog07} because their proofs are not so easy to understand at least for a tiny little old mathematician.

First of all, we prepare
\begin{lemma}\label{primitiveFubini}
Let $u(x,\theta)=\sum_{|a|\le n}\theta^a u_a(x)$ be supersmooth on ${\mathfrak{U}}={\mathfrak{U}}_{\mathrm{ev}}\times\rod^n$.
If $\displaystyle{\int_{{\mathfrak{U}}_{\mathrm{ev}}}} dx \,u_a(x)$ exists for each $a$,  then
we have
$$
{\Biint}_{\!\!\!\mathfrak{U}} dxd\theta\,u(x,\theta)=\int_{{\mathfrak{U}}_{\mathrm{ev}}} dx\bigg[\int_{\rod^n} d\theta \,u(x,\theta)\bigg]
=\int_{\rod^n} d\theta\bigg[\int_{{\mathfrak{U}}_{\mathrm{ev}}} dx \,u(x,\theta)\bigg].
$$
\end{lemma}
\par{\it Proof}. By the primitive definition of integral, we have
$$
{\Biint}_{\!\!\!\mathfrak{U}} dxd\theta\,u(x,\theta)=\int_{{\mathfrak{U}}_{\mathrm{ev}}} dx\bigg[\int_{\rod^n}  d\theta \,u(x,\theta)\bigg]
=\int_{{\mathfrak{U}}_{\mathrm{ev}}} dx \,u_{\tilde1}(x),
$$
and
$$
\int_{\rod^n}  d\theta\sum_{|a|\le n}\bigg[\int_{{\mathfrak{U}}_{\mathrm{ev}}} dx\, \theta^a\,u_a(x)\bigg]
=\int_{\rod^n}  d\theta\sum_{|a|\le n}\theta^a\bigg[\int_{{\mathfrak{U}}_{\mathrm{ev}}} dx\, u_a(x)\bigg]=\int_{{\mathfrak{U}}_{\mathrm{ev}}} dx\, u_{\tilde1}(x). \qed
$$

(I) Now, we consider a simple case:
Let a linear coordinate change be given by
$$
(x,\theta)=(y,\omega)M,\quad M=\begin{pmatrix}
A&C\\
D&B
\end{pmatrix}.
$$
That is,  
$$
x_i=\sum_{k=1}^m y_kA_{ki}+\sum_{\ell=1}^n \omega_\ell D_{\ell i}
=x_i(y,\omega),\quad 
\theta_j=\sum_{k=1}^m y_k C_{kj}+\sum_{\ell=1}^n \omega_\ell B_{\ell j}
=\theta_j(y,\omega)
$$
with $A_{ki},B_{\ell j}\in\cev$ and $C_{\ell i}, D_{kj}\in\cod$,
and we have 
\begin{equation}
\sdet\bigg(\frac{\partial(x,\theta)}{\partial(y,\omega)}\bigg)=\det A{\cdot}
{\det}^{-1}(B-DA^{-1}C)={\det}(A-CB^{-1}D){\cdot}{\det}^{-1}B={\sdet}M.
\label{sdet}
\end{equation}

Interchanging the order of integration, putting $\omega^{(1)}=\omega B$ and $y^{(1)}=yA$, we get
$$
\begin{aligned}
{\Biint} dyd\omega\,&u(y A+ \omega D,y C+\omega B)
=\int  dy \big[\int d\omega\,u(y A+ \omega D,y C+\omega B)\big]\\%\;(y\to Ay,\, \omega\to B\omega)\\
&=\int dy\big[\int d\omega^{(1)}\,\det B{\cdot} u(y A+ \omega^{(1)} B^{-1} D,y C+\omega^{(1)})\big]\\
&=\int d\omega^{(1)}\,\det B\big[\int dy\,u(y A+ \omega^{(1)} B^{-1} D,y C+\omega^{(1)})\big]\\
&=\int d\omega^{(1)}\,\det B\big[\int dy^{(1)}\,\det A^{-1}{\cdot}u(y^{(1)}+ \omega^{(1)} B^{-1} D,y^{(1)}A^{-1}C+\omega^{(1)})\big],
\end{aligned}
$$
that is, since
$$
\frac{\partial(y,\omega)}{\partial(y^{(1)}, \omega^{(1)})}=\begin{pmatrix}
A^{-1}&0\\
0&B^{-1}
\end{pmatrix},\quad
\sdet\bigg(\frac{\partial(y,\omega)}{\partial(y^{(1)}, \omega^{(1)})}\bigg)=\det A^{-1}{\cdot}\det B,
$$
we have
\begin{equation}
\begin{aligned}
{\Biint} dyd\omega\, &u(y A+ \omega D,y C+\omega B)\\
&=
{\Biint} dy^{(1)} d\omega^{(1)}\,
\sdet\bigg(\frac{\partial(y,\omega)}{\partial(y^{(1)}, \omega^{(1)})}\bigg)
{\cdot}u(y^{(1)}+ \omega^{(1)} B^{-1} D,y^{(1)}A^{-1}C+\omega^{(1)}).
\end{aligned}
\label{LCV1}
\end{equation}

Analogously, using Lemma \ref{primitiveFubini} and by introducing change of variables as
$$
y^{(2)}=y^{(1)},\omega^{(2)}=\omega^{(1)}+y^{(1)}A^{-1}C\Longrightarrow
\sdet\bigg(\frac{\partial(y^{(1)},\omega^{(1)})}{\partial(y^{(2)}, \omega^{(2)})}\bigg)
={\sdet
\begin{pmatrix}
1&-A^{-1}C\\
0&1
\end{pmatrix}}
=1,
$$
we get
\begin{equation}
\begin{aligned}
{\Biint} dy^{(1)}d\omega^{(1)}\, & %\sdet\bigg(\frac{\partial(y,\omega)}{\partial(y^{(1)}, \omega^{(1)})}\bigg) 
u(y^{(1)}+\omega^{(1)}B^{-1}D,y^{(1)}A^{-1}C+\omega^{(1)})\\
&=
{\Biint}dy^{(2)}d\omega^{(2)}\,\sdet\bigg(\frac{\partial(y^{(1)},\omega^{(1)})}{\partial(y^{(2)}, \omega^{(2)})}\bigg)
{\cdot}u(y^{(2)}+(\omega-y^{(2)}A^{-1}C)B^{-1}D,\omega^{(2)}).
\end{aligned}
\label{LCV2}
\end{equation}

Then by
$$
\begin{aligned}
&y^{(3)}=y^{(2)}(1-A^{-1}CB^{-1}D),\, \omega^{(3)}=\omega^{(2)}\\
&\Longrightarrow
\sdet\bigg(\frac{\partial(y^{(2)},\omega^{(2)})}{\partial(y^{(3)}, \omega^{(3)})}\bigg)
={\sdet
\begin{pmatrix}
(1-A^{-1}CB^{-1}D)^{-1}&0\\
0&1
\end{pmatrix}}=\det{}^{-1} (1-A^{-1}CB^{-1}D),
\end{aligned}
$$
we have
\begin{equation}
\begin{aligned}
{\Biint} dy^{(2)}d\omega^{(2)}\,& %\sdet\bigg(\frac{\partial(y^{(1)},\omega^{(1)})}{\partial(y^{(2)}, \omega^{(2)})}\bigg)
u(y^{(2)}+(\omega-y^{(2)}A^{-1}C)B^{-1}D,\omega^{(2)})\\
&= %\det{}^{-1} (1-A^{-1}CB^{-1}D)
{\Biint} dy^{(3)}d\omega^{(3)}\,\sdet\bigg(\frac{\partial(y^{(2)},\omega^{(2)})}{\partial(y^{(3)}, \omega^{(3)})}\bigg)
{\cdot}u(y^{(3)}+\omega^{(3)}B^{-1}D,\omega^{(3)}).
\end{aligned}
\label{LCV3}
\end{equation}

Finally by
$$
x=y^{(3)}+\omega^{(3)}B^{-1}D,\,\theta=\omega^{(3)}\Longrightarrow
\sdet\bigg(\frac{\partial(y^{(3)},\omega^{(3)})}{\partial(x, \theta)}\bigg)
={\sdet
\begin{pmatrix}
1&0\\
-B^{-1}D&1
\end{pmatrix}}=1,
$$
using $\det B\,{\det}^{-1}(A-CB^{-1}D){\cdot}
(\det A\,{\det}^{-1}(B-DA^{-1}C))=1$ from \eqref{sdet},  we have,
\begin{equation}
{\Biint} dyd\omega\,u(y A+ \omega D,y C+\omega B)
=\sdet M^{-1}{\cdot} %\det B\,{\det}^{-1}(A-CB^{-1}D)
{\Biint} dxd\theta\,
\sdet\bigg(\frac{\partial(y^{(3)},\omega^{(3)})}{\partial(x, \theta)}\bigg)
{\cdot}u(x,\theta).\quad/\!/
\label{LCV4}
\end{equation}

\begin{remark}
For the linear change of  variables, it is not necessary to assume the compactness of support for integrand using primitive definition of integration.
\end{remark}
(II) (ii-a) If $H_1$ and $H_2$ are superdiffeomorphisms of open subsets of $\supermn$ with the image of $H_1$ 
equals to the domain of $H_2$, then
$$
\Ber(H_1){\cdot}\Ber(H_2)=\Ber(H_2\circ H_1) \where \Ber(H)(y,\omega)=\sdet J(H)(y,\omega).
$$
Here, for $H(y,\omega)=(x_k(y,\omega),\theta_l(y,\omega)):\supermn\to\supermn$, we put 
$$
J(H)(y,\omega)=\begin{pmatrix}
\frac{\partial x_k(y,\omega)}{\partial y_i}&\frac{\partial \theta_l(y,\omega)}{\partial y_i}\\
\frac{\partial x_k(y,\omega)}{\partial \omega_j}&\frac{\partial \theta_l(y,\omega)}{\partial \omega_j}
\end{pmatrix}=\frac{\partial(x,\theta)}{\partial(y,\omega)}.
$$
(ii-b) Any  superdiffeomorphism of an open subset of $\supermn$ may be decomposed as $H=H_2\circ H_1$ where
\begin{equation}
\left\{
\begin{aligned}
&H_1(y,\omega)=(h_1(y,\omega),\omega)=(\tilde{y},\tilde{\omega})\with h_1:\supermn\to{\fR}^{m|0},\\
&H_2(\tilde{y},\tilde{\omega})=(\tilde{y},h_2(\tilde{y},\tilde{\omega}))\with h_2:\supermn\to{\fR}^{0|n}.
\end{aligned}
\right.
\label{20100413-1}
\end{equation}
\begin{remark}
(i) If $H(y,\omega)=(h_1(y,{\omega}),h_2(y,\omega))$ is given by $h_1(y,\omega)=yA+{\omega} D$ 
and $h_2(y,\omega)=yC+{\omega} B$ as above,
putting $H_1(y,\omega)=(yA+{\omega} D,\omega)=(\tilde{y},{\omega})$ and $H_2(\tilde{y},{\omega})=(\tilde{y},\tilde{y}A^{-1}C+{\omega}(B-DA^{-1}C))$, we have $H=H_2\circ H_1$.
In this case, we rewrite the procedures \eqref{LCV1}--\eqref{LCV4} as 
$$
\begin{aligned}
{\Biint} & dyd\omega\,u(yA+{\omega}D, yC+{\omega}B)\\
&={\Biint} d\tilde{y}d\tilde{\omega}\,\sdet\bigg(\frac{\partial(y,\omega)}{\partial(\tilde{y},{\omega})}\bigg)
{\cdot}u(\tilde{y},(\tilde{y}-{\omega}D)A^{-1}C+{\omega}B)\with \tilde{y}=yA+{\omega}D\\
&=\det A^{-1}{\cdot}{\Biint}  d{x}d{\theta}\,\sdet\bigg(\frac{\partial(\tilde{y},{\omega})}{\partial(x,\theta)}\bigg){\cdot}u(x,\theta)
\with x=\tilde{y},\;\theta=\tilde{y}A^{-1}C+{\omega}(B-DA^{-1}C)\\
&=\det A^{-1}{\cdot}{\det}(B-DA^{-1}C){\cdot}{\Biint} dxd\theta\,u(x,\theta).
\end{aligned}
$$
(ii) Analogously, putting $H_1(y,\omega)=(y,yC+{\omega}B)=(y,{\theta})$ and 
$H_2(y,{\theta})=(y(A-CB^{-1}D)+{\theta}B^{-1}D,{\theta})$, 
we have $H=H_2\circ H_1$, and
$$
\begin{aligned}
{\Biint} & dyd\omega\,u(yA+{\omega}D,yC+{\omega}B)\\
&={\Biint} dyd{\theta}\,\sdet\bigg(\frac{\partial(y,\omega)}{\partial(y,{\theta})}\bigg)
{\cdot}u(y(A-CB^{-1}D)+{\theta}B^{-1}D, {\theta})\with  {\theta}=yC+{\omega}B\\
&=\det B{\cdot}{\Biint}  d{x}d{\theta}\,\sdet\bigg(\frac{\partial(y,{\theta})}{\partial(x,\theta)}\bigg){\cdot}u(x,\theta)
\with x=y(A-CB^{-1}D)+CB^{-1}{\theta}\\
&=\det B{\cdot}{\det}^{-1}(A-CB^{-1}D){\cdot}{\Biint} dxd\theta\,u(x,\theta).
\end{aligned}
$$
\end{remark}
(iii) For any given superdiffeomorphism $H(y,\omega)=(h_1(y,\omega),h_2(y,\omega))$, we put
$$
H_1(y,\omega)=(h_1(y,\omega),\omega)=(\tilde{y},{\omega}).
$$
Moreover, using the inverse function $y=g(\tilde{y},\omega)$ of $\tilde{y}=h_1(y,\omega)$, we put
$\tilde{h}_2(\tilde{y},\omega)=h_2(g(\tilde{y},\omega),\omega)$ and 
$H_2(\tilde{y},{\omega})=(\tilde{y},\tilde{h}_2(\tilde{y},{\omega}))$.
Since $h_2(y,\omega)=\tilde{h}_2(h_1(y,\omega),\omega)$, we have $H=H_2\circ H_1$.
We denote
$h_1(y,\omega)=(h_{1j}(y,\omega))=(h_{11},{\cdots}, h_{1m})$ and
$h_2(y,\omega)=(h_{2\ell}(y,\omega))=(h_{21},{\cdots}, h_{2n})$.
Then, for $k, \ell=1,{\cdots},n$,
$$
\frac{\partial \tilde{h}_{2{\ell}}}{\partial \omega_k}=\frac{\partial h_{2{\ell}}}{\partial \omega_k}
+\sum_{i=1}^m\frac{\partial g_i}{\partial \omega_k}\frac{\partial h_{2{\ell}}}{\partial y_i}
$$
with
$$
0=\frac{\partial \tilde{y}_j}{\partial \omega_k}=\frac{\partial h_{1j}(g(y,\omega),\omega)}{\partial \omega_k}
=\sum_{j=1}^m\frac{\partial g_i}{\partial \omega_k}\frac{\partial h_{1j}}{\partial y_i}+\frac{\partial h_{1j}}{\partial \omega_k},
$$
we get
$$
\frac{\partial \tilde{h}_{2{\ell}}}{\partial \omega_k}=\frac{\partial h_{2{\ell}}}{\partial \omega_k}
-\sum_{i,j=1}^m\frac{\partial h_{1j}}{\partial \omega_k}\bigg(\frac{\partial h_{1j}}{\partial y_i}\bigg)^{-1}\frac{\partial h_{2{\ell}}}{\partial y_i}.
$$

Therefore, %taking the transposition of matrices, we have
$$
\Ber H=\sdet
\begin{pmatrix}
\frac{\partial h_1}{\partial y}&\frac{\partial h_2}{\partial y}\\
\frac{\partial h_1}{\partial \omega}&\frac{\partial h_2}{\partial \omega}
\end{pmatrix}
=\det \frac{\partial h_1}{\partial y}{\cdot}
\det{}^{-1}\bigg(
\frac{\partial h_2}{\partial \omega}
-\frac{\partial h_1}{\partial \omega}\bigg(\frac{\partial h_1}{\partial y}\bigg)^{-1}\frac{\partial h_2}{\partial y}\bigg)
=\det \frac{\partial h_1}{\partial y}{\cdot}\det{}^{-1}\frac{\partial \tilde{h}_2}{\partial \omega}.
$$
(III) For each type of superdiffeomorphisms $H_1$ and $H_2$, we need to prove the formula.\\
(III-1) Let $H(y,\omega)=(h(y,\omega),\omega)$ where $h=(h_j)_{j=1}^m:\supermn\to{\fR}^{m|0}$.
Then it is clear that
$$
\Ber(H)(y,\omega)=\det\bigg(\frac{\partial h_j(y,\omega)}{\partial y_i}\bigg)
=\sum_{\sigma\in{\wp}_m}\sgn(\sigma)\prod_{i=1}^m\frac{\partial h_{\sigma(i)}(y,\omega)}{\partial y_i}.
$$
For any $u(x,\theta)=\sum_{|a|\le n}\theta^a u_a(x)$, % with $f_a(q)$ having compact support,
we put
$$
{\Biint}_{\!\!\!{\mathfrak{U}}}dx d\theta\,u(x,\theta)=\int_{{\mathfrak{U}}_{\mathrm{ev, B}}}dx\bigg(\int_{{\fR}^{0|n}}d\theta\,u(x,\theta)\bigg)
=\int_{{\mathfrak{U}}_{\mathrm{ev, B}}}dx\, u_{\tilde 1}(x).
$$
On the other hand, we have
\begin{equation}
\begin{aligned}
{\Biint}_{\!\!\!\mathfrak{V}}& dyd\omega\,\Ber(H)(y,\omega)(u\circ H)(y,\omega)\\
&=\int_{\pi_{\mathrm{B}}(\mathfrak{V})}dy \,\frac{\partial}{\partial\omega_n}{\cdots}\frac{\partial}{\partial\omega_1}
\bigg(\det\bigg(\frac{\partial h_j(y,\omega)}{\partial y_i}\bigg)u(h(y,\omega),\omega)\bigg)\bigg|_{\omega=0}
=(I)+(II)
\end{aligned}
\label{PCV}
\end{equation}
with
$$
\begin{aligned}
(I)&=\int_{\pi_{\mathrm{B}}(\mathfrak{V})}dy
\bigg(\det\bigg(\frac{\partial h_j(y,0)}{\partial y_i}\bigg) u_{\tilde 1}(h(y,0))\bigg),\\
(II)&=\int_{\pi_{\mathrm{B}}(\mathfrak{V})}dy\,\frac{\partial}{\partial\omega_n}{\cdots}\frac{\partial}{\partial\omega_1}
\bigg(\sum_{|a|<n}\omega^a u_a(h(y,\omega))\det\bigg(\frac{\partial h_j(y,\omega)}{\partial y_i}\bigg)\bigg)\bigg|_{\omega=0}.
\end{aligned}
$$
Applying the standard integration on $\euc^m$ to {\it (I)}, we have readily
$$
\int_{\pi_{\mathrm{B}}(\mathfrak{V})}dy\bigg( \det\bigg(\frac{\partial h_j(y,0)}{\partial y_i}\bigg) u_{\tilde 1}(h(y,0))\bigg)
=\int_{{\mathfrak{U}}_{\mathrm{ev, B}}}dx\,u_{\tilde 1}(x)\where {\mathfrak{U}}=H(\mathfrak{V}).
$$

\begin{claim}
{\it (II)} of \eqref{PCV} 
equals to the total derivatives of even variables.
More precisely, we have, for $u(x,\theta)=\sum_{|a|\le n}\theta^au_a(x)$,
$$
%\begin{aligned}
\frac{\partial}{\partial\omega_n}{\cdots}\frac{\partial}{\partial\omega_1}
\bigg(\sum_{|a|< n}\omega^a u_a(h(y,\omega))\Ber(H)(y,\omega)\bigg)\bigg|_{\omega=0}
%\\&=\sum_{\scriptstyle{|a+b|=n, |a|<n}\atop\scriptstyle{|b|={\mathrm{even}}}}
%(-1)^{\tau(a,b)}\partial_{\omega}^b[ f_a(h(y,\omega))\Ber(H)(y,\omega)]\big|_{\omega=0}
=\sum_{j=1}^m\frac{\partial}{\partial y_j}(*).
%\end{aligned}
$$
\end{claim}

\begin{remark}
Though Rogers gives this claim in one sentence,  line 3 from the bottom of p.142 of ~\cite{rog07}, we give a long and naive proof.
\end{remark}

As $h_j(y,\omega)\in\rev$, we have
$$
\begin{aligned}
&h_j(y,\omega)=h_{j\bar{0}}(y)+\sum_{|c|={\mathrm{ev}}\ge2}\omega^c h_{j,c}(y),\\
&u_a(h(y,\omega))=u_a(h_{\bar{0}}(y))+\sum_{|c|={\mathrm{ev}}\ge2}\omega^c h_{j,c}(y)u_{a,x_j}(h_{\bar{0}}(y))
+\sum_{|\alpha|\ge2}\frac{\partial_x^{\alpha}u_a(h_{\bar{0}}(y))}{\alpha!}(\sum_{|c|={\mathrm{ev}}\ge2}\omega^c h_{j,c}(y))^{\alpha},\\
&
{\begin{aligned}
\Ber(H)(y,\omega)&=\det\bigg(\frac{\partial h_j(y,\omega)}{\partial y_i}\bigg)
=\sum_{\sigma\in{\wp}_m}\sgn(\sigma)\prod_{i=1}^m\frac{\partial h_{\sigma(i)}(y,\omega)}{\partial y_i}\\
&=\det\bigg(\frac{\partial h_{j,\bar{0}}(y)}{\partial y_i}\bigg)
+\sum_{\sigma\in{\wp}_m}\sgn(\sigma)
\sum_{j=1}^m\sum_{|c|={\mathrm{ev}}\ge2}\omega^c \frac{\partial h_{\sigma(j),c}(y)}{\partial y_j}
\prod_{i=1,i\neq j}^m\frac{\partial h_{\sigma(i),\bar{0}}(y)}{\partial y_i}\\
&\qquad
+\sum_{\sigma\in{\wp}_m}\sgn(\sigma)
\sum_{j,k=1}^m\sum_{\scriptstyle{|c_j|={\mathrm{ev}}}\atop\scriptstyle{|c_1+c_2|=|c|\ge 4}}
\omega^c \frac{\partial h_{\sigma(j),c_1}(y)}{\partial y_j}
\frac{\partial h_{\sigma(k),c_2}(y)}{\partial y_k}
\prod_{i=1,i\neq j,k}^m\frac{\partial h_{\sigma(i),\bar{0}}(y)}{\partial y_i}
+\mbox{etc}.
\end{aligned}}
\end{aligned}
$$

%\newpage

Putting $\bar{1}-a=b$ or $=c_1+c_2$, $=c_1+c_2+c_3$, etc, we have
\begin{equation}
\mbox{the coefficient of $\omega^b$ of}\, u_a(h(y,\omega))\Ber(H)(y,\omega)={\mathrm{I}}+{\mathrm{II}}+{\mathrm{III}}
\label{100112}
\end{equation}
where
$$
\begin{aligned}
{\mathrm{I}}&=\sum_{j=1}^m h_{j,b}(y)u_{a,x_j}(h_{\bar{0}}(y))\sum_{\sigma\in{\wp}_m}\sgn(\sigma)\prod_{i=1}^m\frac{\partial h_{\sigma(i),\bar{0}}(y)}{\partial y_i},\\
{\mathrm{II}}&=u_a(h_{\bar{0}}(y))
\sum_{\sigma\in{\wp}_m}\sgn(\sigma)
\sum_{j=1}^m \frac{\partial h_{\sigma(j),b}(y)}{\partial y_j}
\prod_{i=1,i\neq j}^m\frac{\partial h_{\sigma(i),\bar{0}}(y)}{\partial y_i},\\
{\mathrm{III}}&=u_a(h_{\bar{0}}(y))
\sum_{\sigma\in{\wp}_m}\sgn(\sigma)
\sum_{j,k=1}^m \sum_{b=c_1+c_2}\frac{\partial h_{\sigma(j),c_1}(y)}{\partial y_j}\frac{\partial h_{\sigma(k),c_2}(y)}{\partial y_k}
\prod_{i=1,i\neq j,k}^m\frac{\partial h_{\sigma(i),\bar{0}}(y)}{\partial y_i}+\mbox{etc}.
\end{aligned}
$$
The term ${\mathrm{II}}$ is calculated as
$$
{\mathrm{II}}=
\sum_{j=1}^m \frac{\partial}{\partial y_j}\bigg[u_a(h_{\bar{0}}(y))
\sum_{\sigma\in{\wp}_m}\sgn(\sigma)
h_{\sigma(j),b}(y)\prod_{i=1,i\neq j}^m\frac{\partial h_{\sigma(i),\bar{0}}(y)}{\partial y_i}\bigg]-A-B
$$
where
$$
\begin{aligned}
A&=
\sum_{j=1}^m \bigg(\sum_{k=1}^m\frac{\partial h_{k,\bar{0}}(y)}{\partial y_j}u_{a,x_k}(h_{\bar{0}}(y))\bigg)
\sum_{\sigma\in{\wp}_m}\sgn(\sigma)h_{\sigma(j),b}(y)\prod_{i=1,i\neq j}^m\frac{\partial h_{\sigma(i),\bar{0}}(y)}{\partial y_i},\\
B&=
\sum_{j=1}^m u_a(h_{\bar{0}}(y))
\sum_{\sigma\in{\wp}_m}\sgn(\sigma)
h_{\sigma(j),b}(y)\frac{\partial}{\partial y_j}\bigg(\prod_{i=1, i\neq j}^m\frac{\partial h_{\sigma(i),\bar{0}}(y)}{\partial y_i}\bigg).
\end{aligned}
$$

Now, we want to prove
(i) $A={\mathrm{I}}$, (ii) $B=0$ and (iii) ${\mathrm{III}}=0$.

(i) To prove $A={\mathrm{I}}$, for each $k=1,{\cdots},m$, we take all sums w.r.t. $\sigma\in{\wp}_m$ and $j$ such that $\sigma(j)=k$. Then, relabeling in $A$, we have
$$
\begin{aligned}
\sum_{\sigma\in{\wp}_m}\sum_{j=1}^m&
\frac{\partial h_{\sigma(j),\bar{0}}(y)}{\partial y_j}
u_{a,x_k}(h_{\bar{0}}(y))\sgn(\sigma)h_{k,b}(y)\prod_{i=1,i\neq j}^m\frac{\partial h_{\sigma(i),\bar{0}}(y)}{\partial y_i}\\
&\qquad\qquad\qquad\qquad
=u_{a,x_k}(h_{\bar{0}}(y))h_{k,b}(y)
\sum_{\sigma\in{\wp}_m}\sgn(\sigma)\prod_{i=1}^m\frac{\partial h_{\sigma(i),\bar{0}}(y)}{\partial y_i}.
\end{aligned}
$$
(ii) Take two permutations $\sigma$ and $\tilde{\sigma}$ in $\wp_m$ such that
$$
\sigma(i)=\tilde{\sigma}(j),\;\; \sigma(j)=\tilde{\sigma}(i),\;\;,\sigma(k)=\tilde{\sigma}(k) \for k\neq i, j,\et \sgn(\sigma)\sgn(\tilde{\sigma})=-1.
$$
Then,
$$
\sgn(\sigma) h_{\sigma(j),b}(y)\frac{\partial}{\partial y_j}\bigg(\prod_{i=1, i\neq j}^m\frac{\partial h_{\sigma(i),\bar{0}}(y)}{\partial y_i}\bigg)
+\sgn(\tilde{\sigma}) h_{\tilde{\sigma},b}(y)\frac{\partial}{\partial y_j}\bigg(\prod_{i=1, i\neq j}^m\frac{\partial h_{\tilde{\sigma}(i),\bar{0}}(y)}{\partial y_i}\bigg)
=0.
$$
(iii) Interchanging the role of $j$, $k$ and $c_1$, $c_2$ in ${\mathrm{III}}$, we have ${\mathrm{III}}=0$.
Others are treated analogously.

Therefore,
$$
{\mathrm{I}}+{\mathrm{II}}+{\mathrm{III}}=A+B=\sum_{j=1}^m \frac{\partial}{\partial y_j} (*)
$$
and we have proved the claim above. /\!/

\begin{corollary} If we assume the compactness of the support of $u_a(x)$ for $|a|<n$,  then we get
$$
\int_{\pi_{\mathrm{B}}(\mathfrak{U})}dy \frac{\partial}{\partial y_i}\big(u_{a}(h(y,\omega))\partial_{\omega}^{\bar{1}-a}\Ber(H)(y,\omega)\big)\big|_{\omega=0}=0.
$$
\end{corollary}

(III-2) For $H(y,\omega)=(y,\phi(y,\omega))$ with $\phi(y,\omega)=(\phi_1(y,\omega),{\cdots},\phi_n(y,\omega))\in{\fR}^{0|n}$, 
we may claim
\begin{equation}
{\Biint}_{\mathfrak{V}}dxd\theta \,u(x,\theta)={\Biint}_{\mathfrak{U}}dyd\omega \bigg(\det\bigg(\frac{\partial\phi_i}{\partial\omega_j}\bigg)\bigg)^{-1}u(y,\phi(y,\omega)).
\label{R11.37}
\end{equation}
In fact, by the analogous proof in (6) of Proposition \ref{4.6}, i.e. odd change of variables formula,
we have the above readily. $\qed$

\subsection{Modification of Vladimirov and Volovich's approach }
We need to check the well-definedness of Definition \ref{4.9} given in the introduction.
First of all, we remark that
by the algebraic nature of integral w.r.t. odd variables, we may interchange the order of integration as
$$
\begin{aligned}
\int_{\rod^n} d{\vartheta}\,\bigg[\int_{\Omega} dq\,\sdet J(\gamma)(q,{\vartheta}){\cdot}u(\gamma(q,{\vartheta}))\bigg]
&=\frac{\partial}{\partial{\vartheta}_n}{\cdots}\frac{\partial}{\partial{\vartheta}_1}\int_{\Omega} dq\,\sdet J(\gamma)(q,{\vartheta}){\cdot}u(\gamma(q,{\vartheta}))\bigg|_{{\vartheta}=0}\\
&=\int_{\Omega} dq\,\frac{\partial}{\partial{\vartheta}_n}{\cdots}\frac{\partial}{\partial{\vartheta}_1}\big(\sdet J(\gamma)(q,{\vartheta}){\cdot}u(\gamma(q,{\vartheta}))\big)\bigg|_{{\vartheta}=0}\\
&=\int_{\Omega} dq\,\bigg[\int_{\rod^n} d{\vartheta}\,\sdet J(\gamma)(q,{\vartheta}){\cdot}u(\gamma(q,{\vartheta}))\bigg]. 
\end{aligned}
$$
In case when $\gamma_{\bar0}(q,{\vartheta})$ doesn't depend on ${\vartheta}$,
putting $\bar{\vartheta}=\gamma_{\bar1}(q,{\vartheta})$ and $\bar q=\gamma_{\bar0}(q)$, we have
$$
\begin{aligned}
\int_{\Omega} dq\,&\frac{\partial}{\partial{\vartheta}_n}{\cdots}\frac{\partial}{\partial{\vartheta}_1}\big(\sdet J(\gamma)(q,{\vartheta}){\cdot}u(\gamma(q,{\vartheta}))\big)\bigg|_{{\vartheta}=0}\\
&=\int_{\Omega} dq\,\det\bigg(\frac{\partial \gamma_{\bar0}(q)}{\partial q}\bigg)
\bigg[\int_{\rod^n} d{\vartheta}\,\det{}^{-1}\bigg(\frac{\partial \gamma_{\bar1}(q,{\vartheta})}{\partial{\vartheta}}\bigg){\cdot}u(\gamma_{\bar0}(q), \gamma_{\bar1}(q,{\vartheta}))\bigg]\\
&=\int_{\Omega} dq\,\det\bigg(\frac{\partial \gamma_{\bar0}(q)}{\partial q}\bigg)
\bigg[\int_{\rod^n} d\bar{{\vartheta}}\,u(\gamma_{\bar0}(q),\bar{{\vartheta}})\bigg]\\
&=\int d\bar{q}\int d\bar{{\vartheta}}\, u(\bar{q},\bar{{\vartheta}})
=\int_{\gamma_{\bar0}(\Omega)}dx\bigg[\int_{\rod^n} d\theta\, u(x,\theta)\bigg]. 
\end{aligned}
$$
That is, for ${\mathfrak{M}}=\gamma(\tilde\Omega)$ with $\tilde\Omega=\Omega\times\rod^n$, $\gamma(q,{\vartheta})=(\gamma_{\bar0}(q),\gamma_{\bar1}(q,{\vartheta}))$,
\begin{equation}
\begin{aligned}
{\VViint}_{\mathfrak{M}} dxd\theta\,u(x,\theta)
&=\int_{\rod^n} d\theta\,\bigg[\int_{\gamma_{\bar0}(\Omega)}dx\, u(x,\theta)\bigg]=\int d\theta\bigg(\int dx \,u(x,\theta)\bigg)\\
&=\int_{\gamma_{\bar0}(\Omega)} dx\,\bigg[\int_{\rod^n} d\theta\,u(x,\theta)\bigg]
=\int dx\bigg(\int d\theta \,u(x,\theta)\bigg).
\end{aligned}
\label{VV4.1}
\end{equation}

Moreover, we need to verify %to prepare
\begin{proposition}[Reparametrization invariance]
Let $\Omega$ and $\Omega'$ be domains in $\euc^m$ and we put $\tilde\Omega$ and $\tilde\Omega'$ as above.
We assume $\tilde\Omega$ and $\tilde\Omega'$ are superdiffeomorphic each other, that is, there exist a diffeomorphism
$\phi_{\bar0}:\Omega'\to \Omega$ such that $\frac{\partial\phi_{\bar0}(q')}{\partial q'}$ which is continuous in $\Omega'$ and $\det(\frac{\partial\phi_{\bar0}(q')}{\partial q'})>0$ and a map
$\phi_{\bar1}:\Omega'\times \rod^n\ni(q',{\vartheta}')\to\phi_{\bar1}(q',{\vartheta}')\in\rod^n$ which is supersmooth w.r.t. ${\vartheta}'$ with $\det(\frac{\partial\phi_{\bar1}(q',{\vartheta}')}{\partial{\vartheta}'})\neq0$.
Put
$$
{\mathfrak{M}}'=\{X'=(x',\theta')\;|\; X'=\gamma\circ\phi(q',{\vartheta}'),\;\; (q',{\vartheta}')\in \tilde\Omega'\}\where
\phi(q',{\vartheta}')=(\phi_{\bar0}(q'),\phi_{\bar1}(q',{\vartheta}')).
$$
For a given path $\gamma:\tilde\Omega\to{\fR}^{m|n}$, we define a path $\gamma\circ\phi:\tilde\Omega'\to{\fR}^{m|n}$.
Then, we have
$$
\VViint_{\!\!\!\gamma(\tilde\Omega)} dxd\theta\,u(x,\theta)
={\VViint}_{\!\!\!\gamma\circ\phi(\tilde\Omega')}dx'd\theta'\,u(x',\theta').
$$
\end{proposition}
\par{\it Proof}. By definition, we have
$$
\VViint_{\!\!\!\gamma(\tilde\Omega)} dxd\theta\,u(x,\theta)
=\int_{\rod^n}d{\vartheta}\,\bigg(\int_{\Omega}dq\,\sdet J(\gamma)(q,{\vartheta})
{\cdot}u(\gamma(q,{\vartheta}))\bigg)
$$
and
$$
{\VViint}_{\!\!\!\gamma\circ\phi(\tilde\Omega')}dx'd\theta'\,u(x',\theta')
=\int_{\rod^n}d{\vartheta}'\,\bigg(\int_{\Omega'}dq'\,
\sdet J(\gamma\circ\phi)(q',{\vartheta}'){\cdot}
{\cdot}u(\gamma\circ\phi(q',{\vartheta}'))\bigg).
$$
Using 
$$
\begin{gathered}
\gamma\circ\phi(q',{\vartheta}')
=(\gamma_{\bar0}(\phi_{\bar0}(q'),\phi_{\bar1}(q',{\vartheta}')), \gamma_{\bar1}(\phi_{\bar0}(q'),\phi_{\bar1}(q',{\vartheta}'))),\\
J(\gamma\circ\phi)(q',{\vartheta}')=J(\gamma)(\phi(q',{\vartheta}'){\cdot}J(\phi)(q',{\vartheta}'),\\
\sdet J(\phi)(q',{\vartheta}')=\det{}^{-1}\bigg(\frac{\partial\phi_{\bar1}(q',{\vartheta}')}{\partial{\vartheta}'}\bigg){\cdot}\det\bigg(\frac{\partial\phi_{\bar0}(q')}{\partial q'}\bigg),
\end{gathered}
$$
we have
$$
\sdet J(\gamma\circ\phi)(q',{\vartheta}')
=\det\bigg(\frac{\partial\phi_{\bar0}(q')}{\partial q'}\bigg){\cdot}\det{}^{-1}\bigg(\frac{\partial\phi_{\bar1}(q',{\vartheta}')}{\partial{\vartheta}'}\bigg)
{\cdot}\sdet J(\gamma)(q,{\vartheta})\bigg|{}_{\scriptstyle{q=\phi_{\bar0}(q')}\atop\scriptstyle{{\vartheta}=\phi_{\bar1}(q',{\vartheta}')}}.
$$

Remarking the order of integration, we have
$$
\begin{aligned}
\int_{\rod^n}d{\vartheta}'\,&\bigg(\int_{\Omega'}dq'\,\sdet J(\gamma\circ\phi)(q',{\vartheta}'){\cdot}u(\gamma\circ\phi(q',{\vartheta}'))\bigg)\\
&=\int_{\Omega'}dq'\,\det\bigg(\frac{\partial\phi_{\bar0}(q')}{\partial q'}\bigg)
%\\&\qquad\times
\bigg[\int_{\rod^n}d{\vartheta}'\,
\det{}^{-1}\bigg(\frac{\partial\phi_{\bar1}(q',{\vartheta}')}{\partial{\vartheta}'}\bigg)\big[\sdet J(\gamma)(q,{\vartheta}){\cdot}u(\gamma(q,{\vartheta}))\big]
\bigg|{}_{\scriptstyle{q=\phi_{\bar0}(q')}\atop\scriptstyle{{\vartheta}=\phi_{\bar1}(q',{\vartheta}')}}\bigg]\\
&=
\int_{\Omega}dq\,
\bigg[\int_{\rod^n}d{\vartheta}\,\sdet J(\gamma)(q,{\vartheta}){\cdot}u(\gamma(q,{\vartheta}))\bigg]
%\\&
={\int{\!\!\!\!\!}\int}_{\!\!\!\tilde\Omega}dqd{\vartheta}\,\sdet J(\gamma)(q,{\vartheta}){\cdot}u(\gamma(q,{\vartheta}))\bigg].  \qed
\end{aligned}
$$

\subsubsection{Proof of Theorem \ref{CVF-VV} -- change of variable formula under integral sign}
$$
\begin{CD}
(\tilde\Omega,dqd{\vartheta}) @>\gamma>>({\mathfrak{M}},dxd\theta)@>{u(x,\theta)}>>
{\mathfrak{V}\!\mathfrak{V}}\!\!\!-\!\!{\displaystyle{\int{\!\!\!\!\!}\int}_{\!\!\mathfrak{M}}} dxd\theta\,u(x,\theta)\in {\fR}\\
%\VViint
@|@A{\varphi}A{}A @|{}\\
(\tilde\Omega,dqd{\vartheta}) @>>\delta>({\mathfrak{N}},dyd\omega)@>>\varphi^*u(y,\omega)>
{\mathfrak{V}\!\mathfrak{V}}\!\!\!-\!\!{\displaystyle{{\int{\!\!\!\!\!}\int}_{\!\!\!\varphi^{-1}({\mathfrak{M}})}}}dy d\omega\,{\sdet}J(\varphi)(y,\omega){\cdot}u(\varphi(y,\omega))\in {\fR}.
\end{CD}
$$

By definition, we have paths
$$
\begin{gathered}
\Omega\times\rod^n\ni(q,{\vartheta})\to\gamma(q,{\vartheta})=(x,\theta),\\
\Omega\times\rod^n\ni(q,{\vartheta})\to\gamma_1(q,{\vartheta})=(y,\omega),
\end{gathered}
$$
which are related each other
$$
(x,\theta)=\gamma(q,{\vartheta})=\varphi(y,\omega)=\varphi(\gamma_1(q,{\vartheta})),\quad \gamma_1=\varphi^{-1}\circ\gamma.
$$

\begin{claim} Denoting the pull-back of a ``superform''  as
$$
{\mathfrak{v}}=dx d\theta\, u(x,\theta)\to \varphi^*{\mathfrak{v}}=dy d\omega \,\sdet J(\varphi)(y,\omega){\cdot}
u(\varphi(y,\omega)),
$$
we have
\begin{equation}
{\VViint}_{\!\!\!\varphi^{-1}\circ\gamma(\tilde{\Omega})}\varphi^*{\mathfrak{v}}={\VViint}_{\!\!\!\gamma(\tilde{\Omega})}{\mathfrak{v}}.
\label{10-2-25-4.9}
\end{equation}
\end{claim}
\par
{\it {Proof}}. Since $J(\varphi^{-1}\circ\gamma)=J(\gamma){\cdot}J(\varphi^{-1})$ which yields
$$
\sdet J(\varphi^{-1}\circ\gamma)(q,{\vartheta})(\sdet J(\varphi)(y,\omega)\bigg|_{(y,\omega)=\varphi^{-1}\circ\gamma(q,{\vartheta})}
=\sdet J(\gamma)(q,{\vartheta}),
$$
and by the definitions of path and integral, we have
$$
\begin{aligned}
{\VViint}_{\!\!\!\varphi^{-1}\circ\gamma(\tilde{\Omega})}{\varphi^*}{\mathfrak{v}}
&=\int_{\rod^n}d{\vartheta}
\bigg[\int_{\Omega}dq\, \sdet J({\varphi^{-1}}\circ\gamma)(q,{\vartheta})(\sdet J(\varphi)(y,\omega){\cdot}u(\varphi(y,\omega))\\
\bigg|_{(y,\omega)=\varphi^{-1}\circ\gamma(q,{\vartheta})}\bigg]\\
&=\int_{\rod^n}d{\vartheta}
\bigg[\int_{\Omega}dq\, \sdet J(\gamma)(q,{\vartheta}){\cdot}u(\gamma(q,{\vartheta}))\bigg]
={\VViint}_{\!\!\!\gamma(\tilde{\Omega})}{\mathfrak{v}}.
\end{aligned}
$$
we have the claim. $\qquad\square$

Now, we interpret  \eqref{10-2-25-4.9} as change of variables:
Since we may denote integrals as
$$
{\VViint}_{\!\!\!\gamma(\tilde{\Omega})}{\mathfrak{v}}={\VViint}_{\mathfrak{M}} dxd\theta\, u(x,\theta),
$$
and
$$
{\VViint}_{\!\!\!\varphi^{-1}\circ\gamma(\tilde{\Omega})}\varphi^*{\mathfrak{v}}
={\VViint}_{\!\!\!\varphi^{-1}\mathfrak{M}} dyd\omega\,\sdet J(\varphi)(y,\omega){\cdot}u(\varphi(y,\omega)),
$$
we have
$$
{\VViint}_{\!\!\!\mathfrak{M}} dxd\theta\, u(x,\theta)
={\VViint}_{\!\!\!\varphi^{-1}\mathfrak{M}} dyd\omega\,\sdet J(\varphi)(y,\omega){\cdot}u(\varphi(y,\omega)). \qed
$$

\end{document}